\documentstyle[11pt,epsf]{article}
\input epsf.tex
 1
 1
 1

\def\lsim{\mathrel{\raise.3ex\hbox{$<$\kern-.75em\lower1ex\hbox{$\sim$}}}}
\def\gsim{\mathrel{\raise.3ex\hbox{$>$\kern-.75em\lower1ex\hbox{$\sim$}}}}
\catcode`\@=11
\def\slash{\mathpalette\make@slash}
\def\make@slash#1#2{\setbox\z@\hbox{$#1#2$}%
  \hbox to 0pt{\hss$#1/$\hss\kern-\wd0}\box0}
\catcode`\@=12 

\begin{document}
\noindent
\thispagestyle{empty}
\renewcommand{\thefootnote}{\fnsymbol{footnote}}
\begin{flushright}
{ CERN-TH/99-415}\\
{ hep-ph/0001286}\\
{ January 2000}\\
\end{flushright}
\vspace{.5cm}
\begin{center}
  \begin{Large}\bf
Top--Antitop Pair Production  close to Threshold:\\[0.2cm]
Synopsis of recent NNLO Results\footnote{
Contribution to the Proceedings of the Workshop on ``Physics Studies for a
Future Linear Collider'', Top Quark Working Group.}
\end{Large}
  \vspace{1.5cm}

\begin{large}
 A.~H.~Hoang$^{a}$,
 M.~Beneke$^{a,b}$,
 K.~Melnikov$^{c}$,
 T.~Nagano$^{d}$,
 A.~Ota$^{d}$,
 A.~A.~Penin$^{e}$,
 A.~A.~Pivovarov$^{f}$,
 A.~Signer$^{g}$,
 V.~A.~Smirnov$^{h}$,
 Y.~Sumino$^{d}$,
 T.~Teubner$^{b}$,
 O.~Yakovlev$^{i}$,
 A.~Yelkhovsky$^{j}$
\end{large}

\vspace{1.5cm}
\begin{it}
${}^a$ Theory Division, CERN, CH-1211 Geneva 23, Switzerland\\[.0cm]
${}^b$ Institut f\"ur Theoretische Physik E, RWTH Aachen, 
       D-52056 Aachen, Germany\\[.0cm]
${}^c$ Stanford Linear Accelerator Center, Stanford University,
       Stanford CA 94309, USA\\[0.cm]
${}^d$ Department of Physics, Tohoku University, Sendai, 
       980-8578 Japan\\[0.cm]
${}^e$ Institute for Theoretical Physics, University Hamburg,
       D-22761 Hamburg, Germany\\[0.cm]
${}^f$ Institute for Nuclear Research of the
       Russian Academy of Sciences, Moscow 117312, Russia\\[0.cm]
${}^g$ Department of Physics, University of Durham, Durham DH1 3LE,
       England\\[0.cm]
${}^h$ Nuclear Physics Institute, Moscow State University,
       119889 Moscow, Russia\\[0.cm]
${}^i$ Randall Laboratory of Physics, University of Michigan, 
       Ann Arbor, Michigan 48109-1120, USA\\[0.cm]
${}^j$ Budker Institute for Nuclear Physics, 630090 Novosibirsk, 
       Russia\\[0.cm]
\end{it}

  \vspace{2.5cm}
  {\bf Abstract}\\
\vspace{0.3cm}

\noindent
\begin{minipage}{14.0cm}
\begin{small}
Using non-relativistic effective theories, new next-to-next-to-leading 
order (NNLO)
QCD corrections to the total $t\bar t$ production cross section at the
Linear Collider have been calculated recently. 
In this article the NNLO calculations of several groups
are compared and the remaining uncertainties are discussed.
The theoretical prospects for an accurate determination of top quark
mass parameters are discussed in detail. An outlook on possible future
improvements is given. 
\\[3mm]
%
\end{small}
\end{minipage}
\end{center}
\setcounter{footnote}{0}
\renewcommand{\thefootnote}{\arabic{footnote}}
\vspace{1.2cm}
%
%
%
\newpage
\noindent
\section{Introduction}
\label{sectionintroduction}
Top--antitop quark pair production close to the threshold
will provide an integral part of the top quark physics program at the
Linear Collider (LC). The theoretical interest in the top--antitop
quark threshold arises from the fact that the large top quark mass 
and width
($\Gamma_t\approx 1.5$~GeV) lead to a suppression of
non-perturbative effects~\cite{Kuehn1,Bigi1,Fadin1}. This makes
perturbative methods a reliable tool to describe the physics of
non-relativistic $t\bar t$ pairs, and allows for measurements of top
quark properties directly at the parton level. Due to the large top
width the total $t\bar t$ production cross section line shape is a
smooth function of the energy, which rises rapidly at the point where
the remnant of a toponium 1S resonance can be formed.
From the energy where this increase occurs, the top quark mass can be
determined, whereas shape and height of the cross section near 
threshold  
can be used to determine $\Gamma_t$, the coupling strength of top
quarks to gluons and, if the Higgs boson is not heavy, the top Yukawa
coupling~\cite{Strassler1}. From differential quantities, such as the
top momentum distribution~\cite{Jezabek1,Suminoold}, 
the forward--backward asymmetry or certain leptonic
distributions~\cite{otherNLO,NLOnonfactordist}, one can obtain
measurements of $\Gamma_t$, the top quark spin and possible anomalous
couplings.   

The measurements of the top quark
mass and the total top quark width from a threshold line shape scan are
particularly interesting.  In contrast to the standard top mass
determination method, which relies on the reconstruction of the
invariant mass of jets originating from 
a single top quark, the line shape measurement  
has the advantage that only colour-singlet $t\bar t$ events have to be
counted. Therefore, the effects of final state interactions 
are suppressed, and systematic uncertainties in the top mass
determination are small. For the total top quark width only a few
other ways to determine it directly are known.
Simulation studies, which also took into account the smearing of the
c.m.~energy from beam effects, have shown that, for a total luminosity
of 100~fb$^{-1}$, statistical and systematical experimental
uncertainties in the top mass determination are below
50~MeV~\cite{Peralta1}. The top quark width can be determined with
experimental uncertainties of better than 20\% for given top quark
mass and $\alpha_s$~\cite{Fujii1,Comas1,Cinabro1}. 

With this prospect in view it is obvious that a careful analysis and
assessment of theoretical uncertainties in the
prediction of the total cross section is mandatory, in order to
determine whether the theoretical precision can meet the experimental
one. Within the last two years, considerable progress has been achieved
in higher order calculations of the total cross section. Using the
concept of effective field theories, calculations of NNLO
QCD corrections to the total cross section have been carried out by
several groups: Hoang--Teubner~\cite{Hoang1,Hoang2},
Melnikov--Yelkhovsky~\cite{Melnikov1},
Yakovlev~\cite{Yakovlev1}, Beneke--Signer--Smirnov~\cite{Beneke1},
Nagano--Ota--Sumino~\cite{Nagano1} and Penin--Pivovarov~\cite{Penin1}. In
contrast to previous LO~\cite{LOttbar} and NLO 
calculations~\cite{Strassler1,Jezabek1,Suminoold,otherNLO,NLOnonfactordist,NLOttbar}, 
the new results at NNLO do not rely on potential models that need
phenomenological input, but represent first-principle QCD
calculations. 
The results are not just some new higher order corrections, but have
led to a number of surprising and important insights.
The NNLO corrections to the location where the cross section rises
and the height of the cross section were found to be much larger than
expected from the 
known NLO calculations. It was suggested that the large corrections to 
the location of the rise are an artifact of the on-shell (pole) mass 
renormalisation~\cite{Beneke2}. Several authors realized that 
the quark pole mass cannot be extracted with an uncertainty
smaller than ${\cal{O}}(\Lambda_{\rm QCD})$ from non-relativistic
heavy quark--antiquark systems~\cite{Hoang3,Beneke2,Uraltsev1}.
New top quark mass definitions were subsequently employed to allow for
a stable extraction of the top quark
mass parameter~\cite{Beneke1,Hoang2}. The remaining 
uncertainties in the normalisation of the 
cross section seem to jeopardise the measurements of the top width,
the top quark coupling to gluons, and the Higgs boson. The results
obtained by all groups are formally equivalent at the NNLO
level. However, they differ in the use of the
calculational methods and the intermediate regularization
prescriptions, and their treatments of higher order corrections. Apart
from analysing the 
theoretical uncertainties estimated from the result of one individual
group, a comparison of the results obtained from the different groups
serves as an additional useful instrument to assess the theoretical
uncertainties.    
In this article the results for the NNLO
QCD calculations for the total cross section obtained by the individual
groups are compared and an
overview of what has been achieved so far based on the results of all
groups is given. As an outline for possible future work, some remaining open
questions are addressed.

The outline of this note is as follows:
in Sec.~\ref{sectioneffective} a brief introduction into the technical
issues relevant to the calculation of the total cross section at NNLO
is given, and some aspects of the effective field theory approach are
reviewed. In Sec.~\ref{sectionresults} the NNLO QCD calculations
obtained by the 
different groups are compared in the pole mass scheme. 
In Sec.~\ref{sectionmasses} three alternative mass definitions tested
by Beneke--Signer--Smirnov, Hoang--Teubner and Melnikov--Yelkhovsky are
discussed. Section~\ref{sectionopen} contains a brief summary and
mentions some issues that should be addressed in the future.

\section{Total Cross Section at NNLO and Effective Theory
  Approach} 
\label{sectioneffective}
For the total cross section close to the threshold, where the velocity
$v$ of the top quarks is small, $v\ll 1$, the conventional perturbative
expansion in the strong coupling breaks down, owing to singular
terms $\sim(\alpha_s/v)^n$ that arise in the $n$-loop amplitude.
This singularity is caused by the instantaneous Coulomb attraction between
the top quarks, which cannot be treated as a perturbation if their
relative velocity is small. It is therefore
mandatory to resum the terms that are singular in $v$ to all orders in
$\alpha_s$. At LO in the non-relativistic expansion of the
total cross section, this amounts to resumming all terms proportional to
$v(\alpha_s/v)^n$, $n=0,\ldots,\infty$. The most convenient
tool to carry out this resummation is the Schr\"odinger equation
\begin{eqnarray}
\bigg(\,
-\frac{\vec\nabla^2}{M_t^{\rm pole}}  
-\frac{C_F\,\alpha_s}{|\mbox{\boldmath $r$}|}
- (\sqrt{q^2}-2 M_t^{\rm pole})-i \Gamma_t
\,\bigg)\,G(\mbox{\boldmath $r$},\mbox{\boldmath $r$}^\prime,\sqrt{q^2}) & = &
 \delta^{(3)}(\mbox{\boldmath $r$}-\mbox{\boldmath $r$}^\prime) 
\,,
\label{SchroedingerLO}
\end{eqnarray}
where $M_t^{\rm pole}$ and $\Gamma_t$ are the top quark pole mass and
width, respectively. The Schr\"odinger equation has the simple form
shown in Eq.~(\ref{SchroedingerLO}) only in the pole mass scheme. The
decay of the top quark is implemented by adding the term
$i\,\Gamma_t$ to the c.m.~energy $\sqrt{q^2}$~\cite{Fadin1}. At LO in
the non-relativistic expansion, counting $\Gamma_t$ as being of order 
$M_t^{\rm pole}\alpha_s^2$, this is the correct way to implement
electroweak effects~\cite{Beneke1,Hoang2}. The total cross section
$\sigma_{\rm tot}(e^+e^-\to\gamma^*,Z^*\to t\bar t)$ reads 
\begin{eqnarray}
\sigma_{\rm tot}^{\gamma,Z}(q^2) & = &
\sigma_{pt}\, \bigg[\,
Q_t^2 - 
2\,\frac{q^2}{q^2-M_Z^2}\,v_e\,v_t\,Q_t +
\bigg(\frac{q^2}{q^2-M_Z^2}\bigg)^2\,
\Big[ v_e^2+a_e^2 \Big]\,v_t^2
\,\bigg]\,R^v(q^2)
\nonumber \\[2mm] & &
+\,\sigma_{pt}\,\bigg(\frac{q^2}{q^2-M_Z^2}\bigg)^2\,
\Big[ v_e^2 + a_e^2 \Big]\,a_t^2\,R^a(q^2)
\,,
\label{totalcrossfullQCD}
\end{eqnarray}
where
\begin{eqnarray}
\sigma_{pt} & = & \frac{4\,\pi\,\alpha^2}{3\,q^2}
\,,
\\[3mm]
v_f & = & \frac{T_3^f - 2\,Q_f\,\sin^2\theta_W}{2\, \sin\theta_W
  \,\cos\theta_W}
\,,
\\[2mm]
a_f & = & \frac{T_3^f}{2\, \sin\theta_W\, \cos\theta_W}
\,.
\end{eqnarray}
Here, $\alpha$ is the fine structure constant, $Q_t=2/3$ the electric
charge of the top quark, $\theta_W$ the Weinberg angle, and $T_3^f$
refers to the third component of the weak isospin;
$R^v$ and $R^a$ represent the contributions to the cross section
induced by vector- and axial-vector current, respectively.
$R\equiv Q_t^2 R^v$ is equal to the total normalised
photon-induced cross section, which is usually referred to as the
$R$-ratio. Close to threshold, $\sigma_{\rm tot}^{\gamma,Z}$ is dominated 
by the vector-current contribution $R^v$, which describes top quark
pairs in an angular momentum S-wave state. Higher angular momentum states
are suppressed by additional powers of $v$; P-wave production, which
is associated to the axial-vector current contribution $R^a$ is
suppressed by $v^2$ and needs to be taken into account at
NNLO~\cite{Fadin2,Penin1,Hoang2,Kuehn2}. The absorptive part of 
$G(\mbox{\boldmath $0$},\mbox{\boldmath $0$},\sqrt{q^2})$, obtained from
Eq.~(\ref{SchroedingerLO}), is the first term of a non-relativistic
expansion of $R^v$ close to threshold:
\begin{eqnarray}
R^{v}(q^2\approx 4M_t^2) & = &
\frac{72\,\pi}{q^2}\,{\mbox{Im}}\Big[
G(\mbox{\boldmath $0$},\mbox{\boldmath $0$},\sqrt{q^2})\Big] +\ldots
\,.
\end{eqnarray}

To determine NLO corrections to the cross
section, corresponding to a resummation of all terms $\propto
v(\alpha_s/v)^n\times[\alpha_s,v]$, $n=0,\ldots,\infty$, the one-loop
corrections to the Coulomb potential~\cite{Voneloop} 
have to be added in Eq.~(\ref{SchroedingerLO}) and a short-distance
correction to the top--antitop production current has to be
included~\cite{Karplus1}. The latter is implemented by multiplying the
Green function of the Schr\"odinger equation by a factor
$C=1+c_1\frac{\alpha_s}{\pi}$, where $c_1$ is a real number.
The NLO corrections do not pose any conceptual problem, because the
short-distance corrections to $C$ factorise unambiguously, and because
the absorptive part of the Green function does not contain any
ultraviolet divergences at this order. 
The corrections at NNLO, corresponding to a resummation of all terms
$\propto v(\alpha_s/v)^n\times[\alpha_s^2,\alpha_s\,v,v^2]$,
$n=0,\ldots,\infty$, require the inclusion of the kinetic energy 
term $-\frac{\vec\nabla^4}{4\,M_t^3}$, two-loop corrections to the
static potential~\cite{Vtwoloop} and new
potentials suppressed by additional powers of $1/M_t^2$ and
$\alpha_s/M_t$ into Eq.~(\ref{SchroedingerLO}). In addition,
the short-distance factor $C$ has to be determined at 
the two-loop level~\cite{Beneke3,shortdist}. The new potentials are the
generalisation of the 
Breit--Fermi potential, known from positronium, in QCD.
The determination of the NNLO corrections is
non-trivial because the additional mass-suppressed terms
in the Schr\"odinger equation lead to UV divergences in the
absorptive part of the Green function $G$. This is because they
contain momenta to a high positive power. These divergences are a
consequence of the non-relativistic expansion.

The problem of UV divergences can be conveniently dealt
with in the framework of non-relativistic effective
theories.
All the NNLO calculations performed in
Refs.~\cite{Hoang1,Melnikov1,Yakovlev1,Beneke1,Nagano1,Penin1,Hoang2}
have been carried out in this framework. 
In the effective field theory approach the top quark and the gluonic 
degrees of freedom that are off-shell in the non-relativistic
top--antitop-quark system are integrated out, leaving only those
degrees of freedom as dynamical that can become on-shell.
The relevant momentum regimes associated with non-relativistic 
degrees of freedom
have a non-trivial structure~\cite{softliterature} and were finally
identified by Beneke and Smirnov~\cite{Beneke3}. 
The resulting effective field theory obtained by integrating out the
degrees of freedom that are off-shell has been called ``potential
non-relativistic QCD'' (PNRQCD)~\cite{Pineda1}. PNRQCD represents an
effective theory of NRQCD; the latter was first proposed by Caswell and
Lepage~\cite{Caswell1} and is widely used in charmonium and
bottomonium physics~\cite{Bodwin1}. As the NRQCD Lagrangian, the PNRQCD
Lagrangian contains an infinite number of operators, where operators of
higher dimension are associated to interactions suppressed by higher
powers in $v$. The corresponding Wilson coefficients can be determined
perturbatively (as a conventional series in $\alpha_s$) by matching
on-shell scattering amplitudes in PNRQCD and full QCD. The main
feature of PNRQCD is the existence of spatially non-local
instantaneous four-quark interactions, which represent an
instantaneous interaction of a quark--antiquark pair separated by some
spatial distance $\mbox{\boldmath $r$}$:
\begin{eqnarray}
{\cal{L}}^{\mbox{\tiny PNRQCD}}_{\tiny\rm non-local} & = &
\int d^3 \mbox{\boldmath $r$} \Big(\psi^\dagger \psi\Big)
  (\mbox{\boldmath $r$})\,V(\mbox{\boldmath $r$})\,
\Big(\chi^\dagger \chi\Big)(0)
\,.
\label{PNRQCDLagrangian}
\end{eqnarray}
Here, $\psi$ and $\chi$ represent the two-component
Pauli spinors describing the top and antitop quarks after the
corresponding small component has been integrated out.
The Wilson coefficients $V(\mbox{\boldmath $r$})$ of these non-local
interactions are generalisations of the concept of the heavy quark
potential. 
We emphasise that these Wilson coefficients are strict short-distance
quantities that can be calculated perturbatively. 
In addition, PNRQCD 
contains the interactions of dynamical gluons (having energies and
momenta of the order of the top quark kinetic energy) with the top
quarks. These dynamical gluons lead to top--antitop quark interactions
that are not only non-local in space but also in time. These
interactions are called ``retardation effects''. External 
electroweak currents, which describe production and annihilation of a
top quark pair are rewritten in terms of PNRQCD currents, providing
a systematic small-velocity-expansion of the corresponding relativistic
current--current correlators. The Wilson
coefficients of the PNRQCD currents contain short-distance information
specific to the corresponding electroweak current producing or
annihilating the top--antitop-quark pair. 
The short-distance factor $C$ is the modulus square of the Wilson
coefficient of the first term in the non-relativistic
expansion of the vector current.
PNRQCD provides so-called ``velocity counting rules'' that
unambiguously state which of the operators have to be taken into
account to describe the quark--antiquark dynamics at a certain parametric
precision. For the description of a non-relativistic $t\bar t$ pair at
NNLO, these rules show that the interactions of dynamical gluons can be 
neglected. The resulting equation of motion for a heavy
quark--antiquark pair has the form of Eq.~(\ref{SchroedingerLO}),
supplemented by corrections up to NNLO. UV divergences in the
calculation of the absorptive part of the Green function are
subtracted and interpreted in the context of a particular 
regularization scheme for PNRQCD. The absorptive 
part of the Green function then contains a dependence on the
regularization scheme parameter. This dependence on the regularization
parameter is cancelled by that of
the two-loop corrections to the short-distance factor $C$. 

We note that, strictly speaking, the new NNLO calculations represent 
true NNLO results only for
the case of a stable top quark. None of the new NNLO calculations
contains a consistent treatment of electroweak effects at NNLO.
All groups took into account the top quark width by adding
$i\,\Gamma_t$ to the c.m.~energy in the Schr\"odinger equation, as
shown in Eq.~(\ref{SchroedingerLO})\footnote{
Some NNLO corrections proportional to the top width have
been determined in Refs.~\cite{Penin1,Hoang2}.
}.
However, we do not expect that the neglected electroweak
corrections will exceed several per cent for the total cross section. 

\section{NNLO Results for the Total Cross Section\\ in the Pole Mass Scheme}
\label{sectionresults}
Six groups have calculated the NNLO QCD corrections to the total cross
section close to threshold: Hoang--Teubner~\cite{Hoang1,Hoang2},
Melnikov--Yelkhovsky~\cite{Melnikov1}, Yakovlev~\cite{Yakovlev1},
Beneke--Signer--Smirnov~\cite{Beneke1}, Nagano--Ota--Sumino~\cite{Nagano1}
and Penin--Pivovarov~\cite{Penin1}. In this section the methods of the
different groups are briefly summarised and differences are pointed
out. The results are compared numerically in the pole mass scheme.  
Because NNLO corrections are only relevant to the 
vector-current-induced total production cross section, we will only
compare the 
normalised photon-induced cross section $R$. Results for the cross
section including also the full Z-exchange contributions can be found
in Refs.~\cite{Kuehn2,Penin1,Hoang2}. The methods for the NNLO results
from the 
groups Melnikov--Yelkhovsky, Yakovlev and Nagano--Ota--Sumino are
identical. Because the numerical results provided by these groups
for this comparison agree with each other to better than one per mille,   
they will be treated as belonging to a single group.

The different groups have used the following methods in their
NNLO calculations:
\begin{itemize}
\item {\bf Hoang--Teubner} (HT)~\cite{Hoang2} have solved the
  NNLO Schr\"odinger equation exactly in momentum space
  representation. As ultraviolet regularization they restricted all
  momenta to be smaller than the cutoff $\Lambda$, which is of the order
  of the top quark mass. The short-distance coefficient
  $C$ was determined by using the ``direct matching
  procedure''~\cite{Hoang4}, where the total cross section in the
  effective field theory is matched to the total 
  cross section in QCD in the limit $\alpha_s\ll v\ll 1$ and for
  $\Gamma_t=0$. The result for the total cross section depends on two
  scales, $\mu_{\rm soft}$, the renormalisation scale of the strong
  coupling in the static potential and the cutoff scale $\Lambda$. 
  The renormalisation scale in the short-distance coefficient $C$ is
  also $\mu_{\rm soft}$. 
  The sensitivity to $\Lambda$ is considerable at LO and has been
  shown to be small at NLO and NNLO.   
\item {\bf Melnikov--Yelkhovsky--Yakovlev--Nagano--Ota--Sumino}
  (MYYNOS)~\cite{Melnikov1,Yakovlev1,Nagano1} solved the  
  NNLO Schr\"odinger equation exactly in coordinate space representation.
  As regularization prescription they determined the Green function
  at a finite distance $r_0$ from the origin and expanded in $r_0$.
  Only logarithms of $r_0$ were kept and inverse powers of $r_0$ were
  discarded. The value of $r_0$ was chosen of the order of the 
  inverse top quark mass. The short-distance
  coefficient $C$ was determined by using the ``direct matching
  procedure''. The result for the total cross section depends
  on three scales, $\mu_{\rm soft}$, the renormalisation scale in the
  static potential, $1/r_0$, the cutoff scale, and 
  $\mu_{\rm hard}$, the renormalisation scale in the short-distance
  coefficient $C$. The scale $\mu_{\rm hard}$ was set equal to the top
  quark mass. The sensitivity to $r_0$ has been shown to be small.  
\item {\bf Penin--Pivovarov} (PP)~\cite{Penin1} solved the
  Schr\"odinger equation perturbatively in coordinate space
  representation. They  
  started from the analytically known solution of the LO Coulomb
  problem and determined NLO and NNLO corrections analytically via
  Rayleigh-Schr\"odinger time-independent perturbation theory. As
  regularization prescription they determined the Green function at
  a finite distance to the origin, discarding all power-like
  divergences. In order to avoid multiple poles in the energy 
  denominators of the Green function, which naturally arise in a
  perturbative determination of the Green function and which lead to
  instabilities in the cross-section shape, PP supplemented
  their calculation by reabsorbing the corrections to the energy
  eigenvalues into single-pole energy denominators. The short-distance
  coefficient $C$ was determined by using the ``direct matching
  procedure''. The result for the total cross section depends
  on three scales, $\mu_{\rm soft}$, the renormalisation scale in the
  static potential, $\mu_{\rm fac}$, the cutoff scale, and 
  $\mu_{\rm hard}$, the renormalisation scale in the short-distance
  coefficient $C$. The scales $\mu_{\rm fac}$ and $\mu_{\rm hard}$
  have been chosen of the order of the top quark mass. The sensitivity
  to variations of $\mu_{\rm fac}$ and $\mu_{\rm hard}$ has been shown
  to be small.   
\item {\bf Beneke--Signer--Smirnov} (BSS)~\cite{Beneke1} solved the 
  Schr\"odinger equation perturbatively using dimensional
  regularization as a regularization prescription. They started from
  the analytically known solution of the LO Coulomb problem and
  determined all corrections analytically via Rayleigh--Schr\"odinger  
  time-independent perturbation theory. At NLO BSS included the second
  iteration of the one-loop corrections to the static potential. The
  short-distance coefficient $C$ was determined by extracting the hard
  momentum contribution in the two-loop amplitude for $\gamma\to t\bar
  t$ close to threshold for $\Gamma_t=0$~\cite{shortdist}, using the
  ``threshold expansion''~\cite{Beneke3}, 
  which is an algorithm to calculate the asymptotic expansion of
  diagrams describing processes involving massive quark--antiquark
  pairs in the kinematic region close to the two-particle
  threshold. In order to avoid the destabilising effects of multiple
  poles in the energy denominators of the Green function, BSS 
  supplemented their result by reabsorbing the corrections to the two
  lowest lying energy eigenvalues into single-pole energy
  denominators. In contrast to all other groups, BSS have not
  implemented the short-distance coefficient $C$ as a global factor,
  but have expanded $C$ together with the non-relativistic corrections
  to the Green function up to NNLO. The result of BSS depends 
  on the scale $\mu_{\rm soft}$ and on the QCD/NRQCD matching scale
  $\mu_{\rm h}$. The dependence on the scale $\mu_{\rm h}$ has been
  shown to be small.     
\end{itemize}

\begin{figure}[t!] 
\begin{center}
\leavevmode
\epsfxsize=3.8cm
\leavevmode
\epsffile[220 580 420 710]{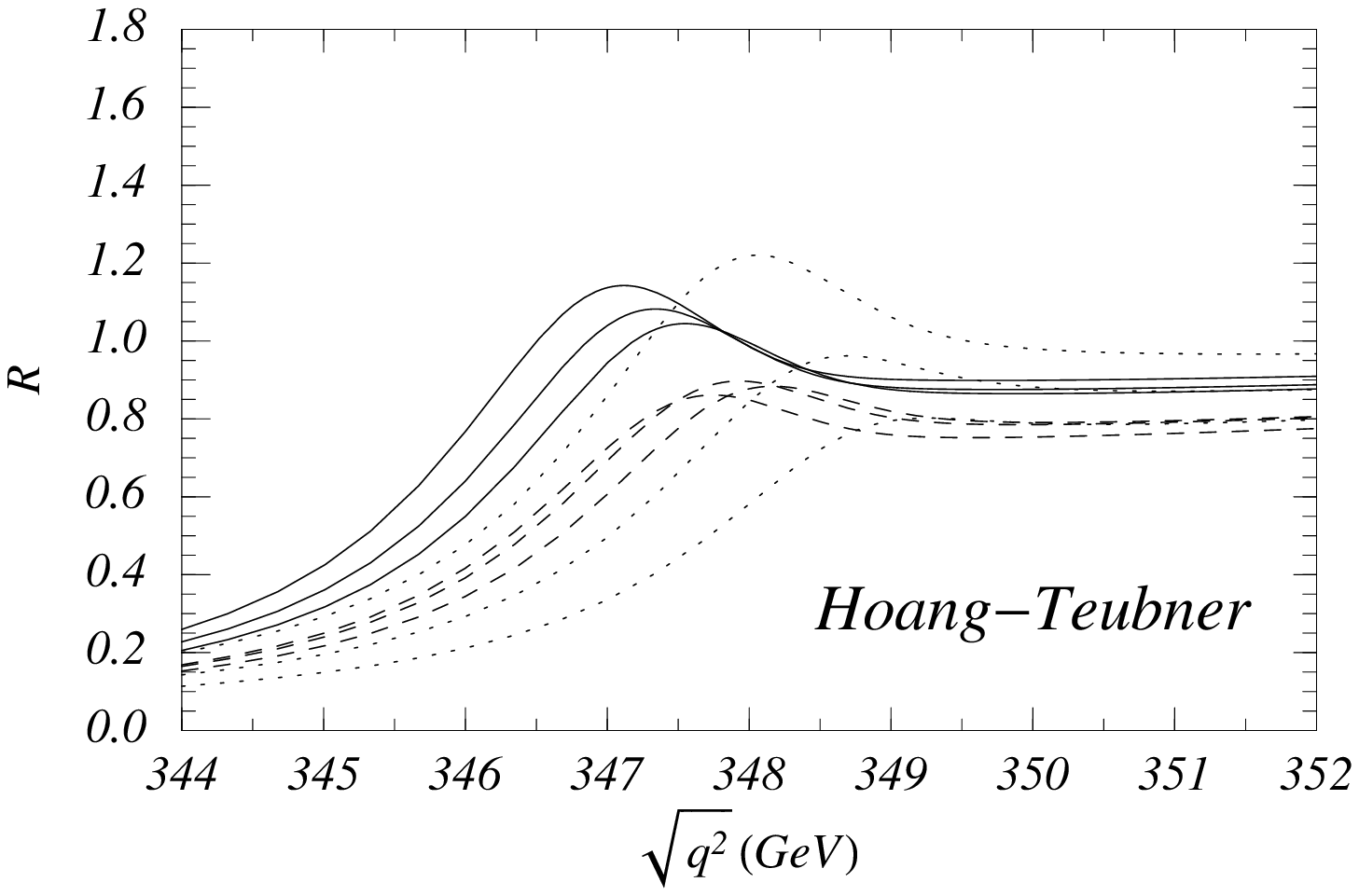}
\hspace{4.1cm}
\epsfxsize=3.8cm
\leavevmode
\epsffile[220 580 420 710]{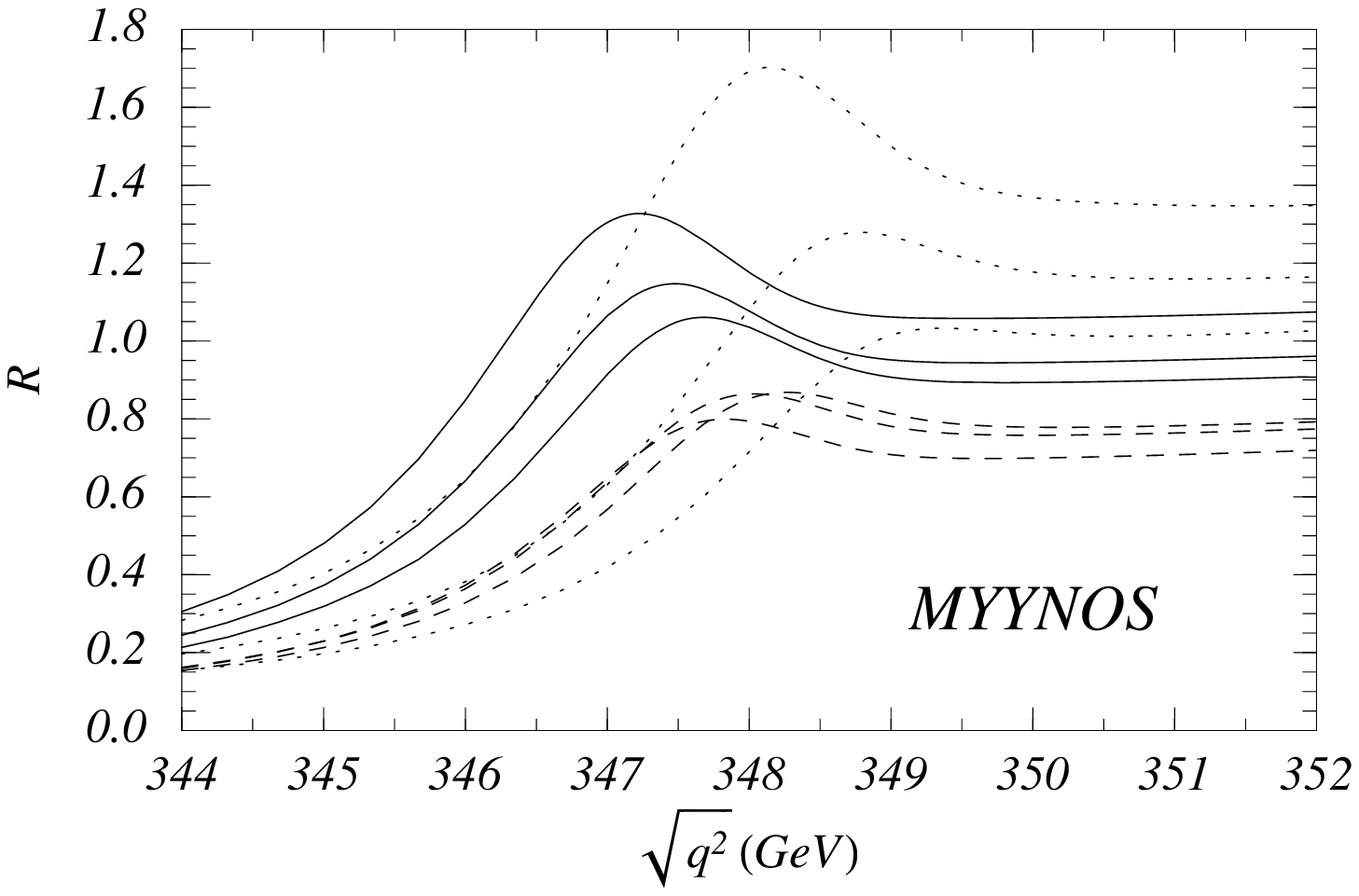}
\vskip 3cm
\leavevmode
\epsfxsize=3.8cm
\leavevmode
\epsffile[220 580 420 710]{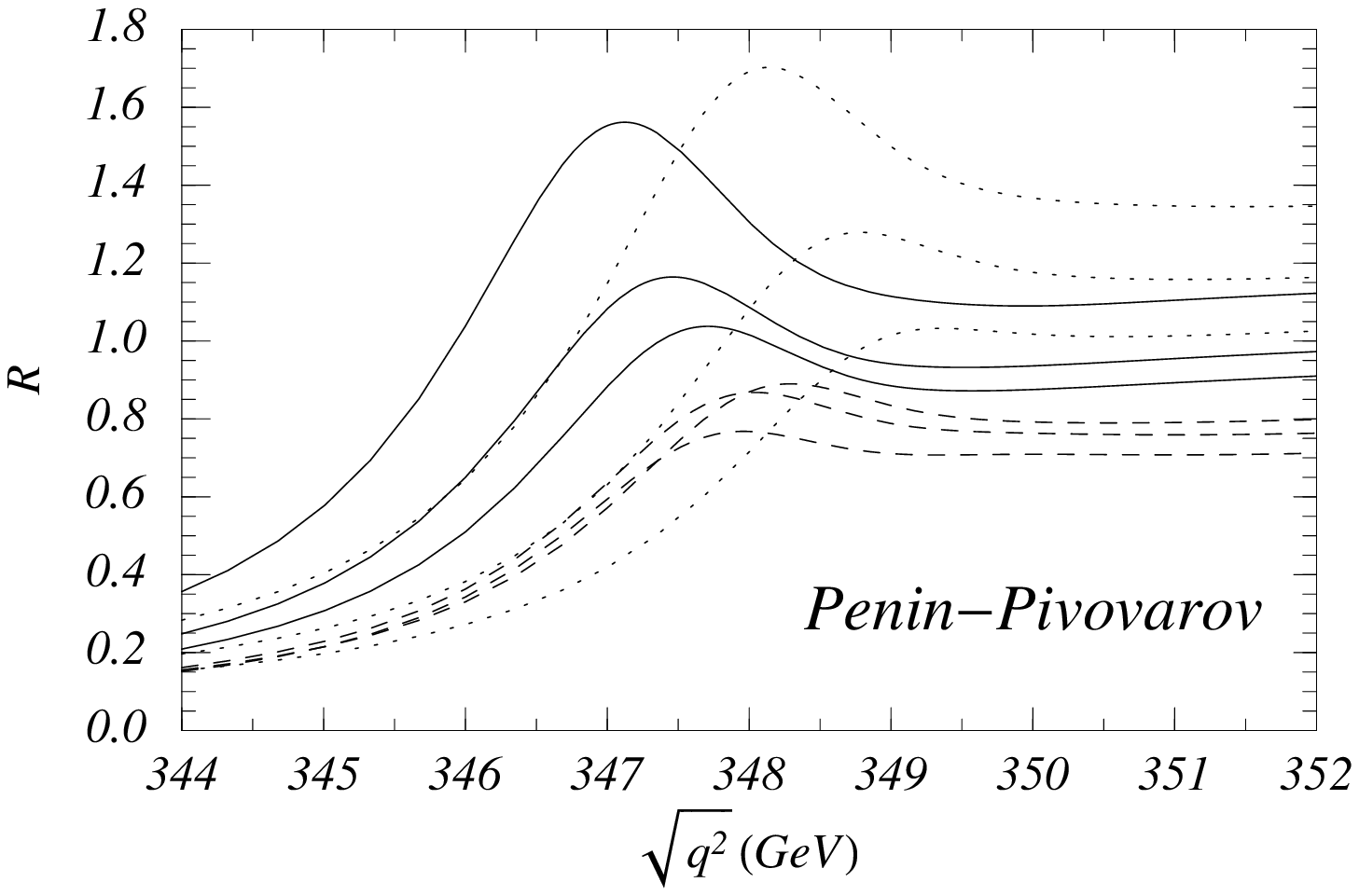}
\hspace{4.1cm}
\epsfxsize=3.8cm
\leavevmode
\epsffile[220 580 420 710]{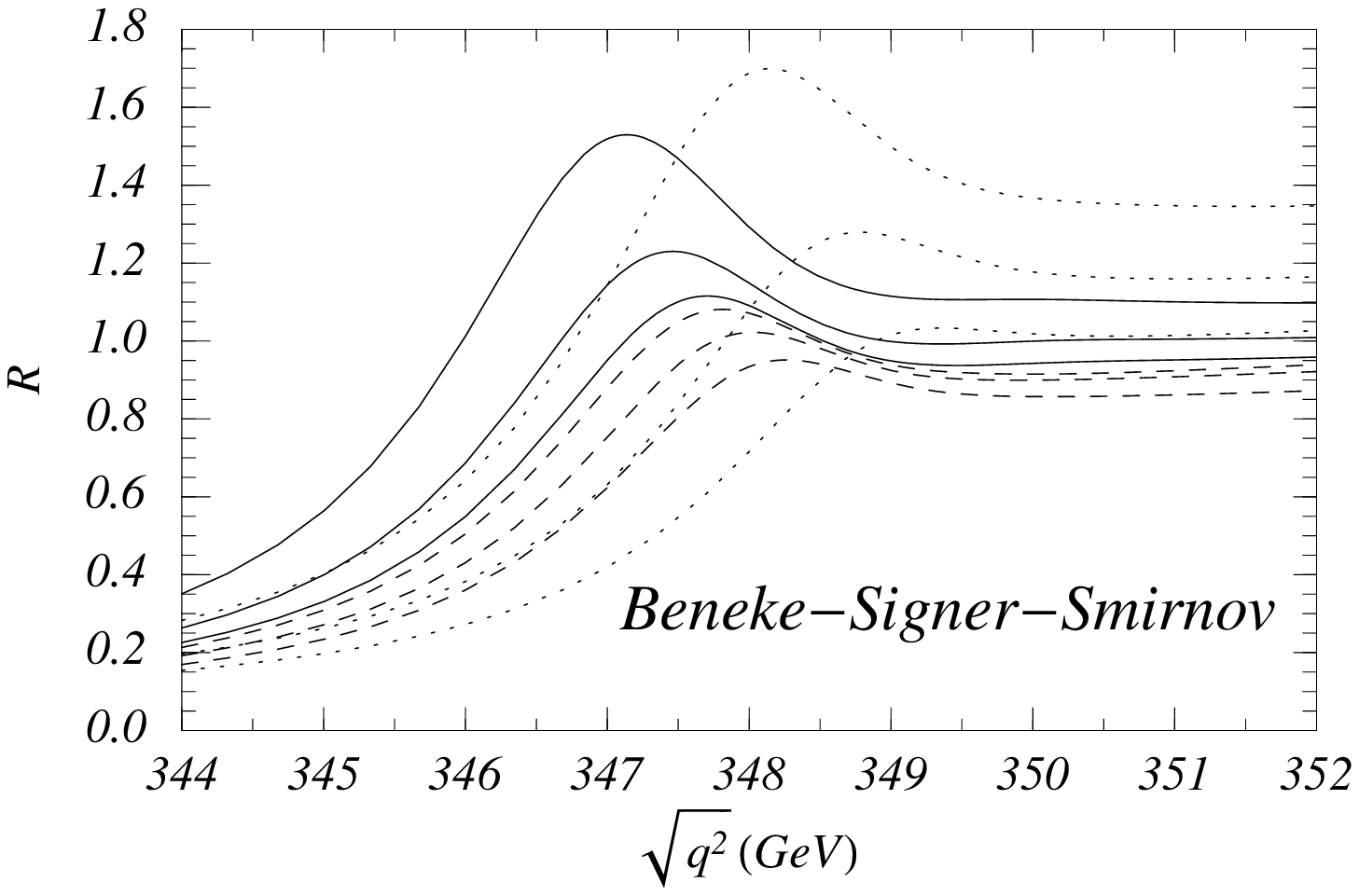}
\vskip  2.5cm
 \caption{\label{figpole}
The total normalised photon-induced $t\bar t$ cross section  $R$ at
the LC versus the c.m. energy in the threshold regime at LO (dotted
curves), NLO (dashed) and NNLO (solid) in the pole mass scheme for
$M_t^{\rm pole}=175.05$~GeV,   
$\alpha_s(M_Z)=0.119$ , $\Gamma_t=1.43$~GeV and $\mu_{\rm soft}=15$,
$30$, $60$~GeV.  
The plots have been generated from results provided by the groups
Hoang-Teubner (HT), Melnikov-Yelkhovsky-Yakovlev-Nagano-Ota-Sumino
(MYYNOS), Penin-Pivovarov (PP) and Beneke-Signer-Smirnov (BSS).
}
 \end{center}
\end{figure}
\begin{table}[t] 
\vskip 7mm
\begin{center}
\begin{tabular}{|c||c|c|c||c|c|c||c|c|c|} \hline
Order & 
\multicolumn{3}{|c||}{LO} 
   & \multicolumn{3}{|c||}{NLO}
   & \multicolumn{3}{|c|}{NNLO} \\ \hline
$\mu_{\rm soft}$[GeV] & $15$ & $30$ & $60$ 
                      & $15$ & $30$ & $60$  
                      & $15$ & $30$ & $60$  \\ \hline
\raisebox{-1.6ex}[1.6ex]{HT}
   & $1.22$ & $0.96$ & $0.80$ 
   & $0.86$ & $0.90$ & $0.88$
   & $1.14$ & $1.08$ & $1.04$ \\[-1mm]
   & $348.06$ & $348.69$ & $349.26$ 
   & $347.75$ & $347.93$ & $348.17$
   & $347.12$ & $347.34$ & $347.55$ \\\hline
\raisebox{-1.6ex}[1.6ex]{MYYNOS}
   & $1.70$ & $1.28$ & $1.03$
   & $0.80$ & $0.86$ & $0.87$
   & $1.33$ & $1.15$ & $1.06$\\[-1mm]
   & $348.14$ & $348.79$ & $349.36$
   & $347.83$ & $348.03$ & $348.27$
   & $347.22$ & $347.48$ & $347.68$ \\\hline
\raisebox{-1.6ex}[1.6ex]{PP}
   & $1.70$ & $1.28$ & $1.03$
   & $0.77$ & $0.87$ & $0.89$
   & $1.56$ & $1.16$ & $1.04$\\[-1mm]
   & $348.14$ & $348.79$ & $349.36$
   & $347.96$ & $348.04$ & $348.29$ 
   & $347.12$ & $347.46$ & $347.71$ \\\hline
\raisebox{-1.6ex}[1.6ex]{BSS}
   & $1.70$ & $1.28$ & $1.03$
   & $1.08$ & $1.02$ & $0.95$
   & $1.53$ & $1.23$ & $1.12$\\[-1mm]
   & $348.15$ & $348.79$ & $349.36$
   & $347.81$ & $348.03$ & $348.26$
   & $347.14$ & $347.46$ & $347.70$ \\\hline
\end{tabular}
\caption{\label{tabpole}
The values of $R$ (upper numbers) at the respective peak position
(lower numbers in units of GeV) at LO, NLO and NNLO in the pole mass
scheme for $M_t^{\rm pole}=175.05$~GeV,  $\alpha_s(M_Z)=0.119$,
$\Gamma_t=1.43$~GeV and 
$\mu_{\rm soft}=15$, $30$, $60$~GeV.  The values have been determined
from results provided by the groups Hoang-Teubner (HT),
Melnikov--Yelkhovsky--Yakovlev--Nagano--Ota--Sumino (MYYNOS),
Penin--Pivovarov (PP) and Beneke--Signer--Smirnov (BSS).  }
\end{center}
\vskip 3mm
\end{table}

In Figs.~\ref{figpole} the total normalised photon-induced cross
section $R$ obtained from HT ($\Lambda=M_t^{\rm pole}$), MYYNOS
($r_0=e^{2-\gamma}/2M_t^{\rm pole}$, $\mu_{\rm hard}=M_t^{\rm pole}$),
PP ($\mu_{\rm fac}=\mu_{\rm hard}=M_t^{\rm pole}$) and BSS
($\mu_{\rm h}=M_t^{\rm pole}$) are displayed at LO (dotted lines), NLO 
(dashed lines) and NNLO (solid lines) in the 
non-relativistic expansion in the pole mass scheme for
$M_t^{\rm pole}=175.05$~GeV, 
$\alpha_s(M_Z)=0.119$, $\Gamma_t=1.43$~GeV 
and $\mu_{\rm soft}=15$, $30$, $60$~GeV.
The value for the top quark pole mass is the highest-order entry of
Table~\ref{tabkinpole}, taking $\overline m_t(\overline m_t)=165$~GeV as
a reference value.
The range $15$--$60$~GeV for $\mu_{\rm soft}$ is chosen, because it
covers the typical top quark three momentum in the $t\bar t$ system.
The effects of the beam energy spread due to initial-state radiation
and beamstrahlung, which lead to a smearing of the effective
centre-of-mass energy and a loss of luminosity, are not included in
this comparison. 
In Table~\ref{tabpole} the values of $R$ (upper numbers) at the
visible maximum (lower numbers in units of GeV) at LO, NLO
and NNLO in the pole mass scheme are displayed using the same set of
parameters as in Figs.~\ref{figpole}. 
The various results presented in Figs.~\ref{figpole} and
Table~\ref{tabpole} 
are of the same order and only differ with respect to the treatment of
higher order corrections, and with respect to the regularization
scheme. 

As far as the position of the maximum, called ``peak position'' in the
rest of this article, is concerned the results of all groups are
consistent: they all show that 
the position of the peak receives large NNLO corrections and that the
peak is moved to smaller c.m. energies at higher orders. 
For $\mu_{\rm soft}=15/30/60$~GeV, the NLO shift is around 
$300/800/1200$~MeV versus $600$--$800/600/600$~MeV at NNLO.
The convergence is better for higher renormalisation scales, but the
size of the overall shift is also increasing. In addition, the
dependence of the peak position on the renormalisation scale is not
reduced when going from NLO to NNLO\footnote{
There is also a rather strong correlation of the peak position to the
choice of $\alpha_s$, which arises from a quadratic dependence of the
peak position on the strong coupling,  
$M_{\rm peak}-2M_t^{\rm pole}=-\frac{4}{9}\alpha_s^2 M_t^{\rm
  pole}[1+\ldots]$. The ellipses denote electroweak and higher order
QCD corrections.  
}. 
At LO, NLO and NNLO, the
variation is around $1200$, $400$ and $400$~MeV respectively. 
An extraction of the top quark pole mass based on the location of the
peak would result in a theoretical uncertainty of around $300$
MeV, although an exact estimate based on the results given above is
difficult. (The uncertainty coming from the use of different
calculational methods by the various groups, for the same input
parameters, is only around 50~MeV at LO and NLO, and around 80~MeV at
NNLO.) The rather bad behaviour of the peak position is not
unexpected, because it is known that the pole mass definition suffers
from a sensitivity to low scales (i.e. scales that are smaller than
the physical scales relevant to the problem), which increases for
higher orders in perturbation theory. This leads to large artificial
corrections in larger orders of perturbation theory. The problem is
known as the ``renormalon problem'' of the pole mass
definition~\cite{renormalon} and exists even in the presence of the
large top quark width. (A formal proof can be found in
Ref.~\cite{Smith1}.) In practice, this means that, as a matter of
principle, the top quark pole mass cannot be determined to better
than ${\cal{O}}(\Lambda_{\rm QCD})$. The
pole mass definition could, 
at least in principle, still be used as a correlated parameter that
would depend on the order of the calculation and the choice of the
theoretical parameters, such as the strong coupling, the
renormalisation scale, etc., but it is wise not to put this option
into practice. A way that avoids the problem of large higher
order corrections to the peak position is to use top quark mass definitions
that do not have the same strong sensitivity to low scales as the pole
mass. Such masses can also be defined in a way that the
correlation of the peak position on the value of the strong coupling
is small. Some suggested alternative mass
definitions~\cite{Beneke2,Hoang2,Bigi2},
called ``threshold masses'' in this article, are discussed in
Sec.~\ref{sectionmasses}.

As far as the normalisation of the cross sections obtained by the
different groups is concerned, all results clearly show
that the sensitivity of the NLO total cross section with respect to 
changes in $\mu_{\rm soft}$ does not give an estimate for the true
size of the NNLO corrections. 
However, for the actual size of the NNLO corrections the 
situation is less coherent. Compared to the other groups, the 
normalisation of the cross sections from
HT has the smallest sensitivity to variations
of $\mu_{\rm soft}$, and the smallest size of NLO and NNLO 
corrections. At the peak position, the value of $R$ from HT varies
by (40,4,10)\% at (LO,NLO,NNLO) for a variation of $\mu_{\rm soft}$ from
$15$ to $60$~GeV, compared to (50,8,23)\% for MYYNOS, (55,14,45)\% for PP and
(52,6,33)\% for BSS. The NLO and NNLO corrections to $R$ at the peak
position for 
$\mu_{\rm soft}=(15,30,60)$~GeV amount to $(0.36,0.06,0.08)$ and
$(0.28,0.18,0.16)$ for HT, 
$(0.90,0.42,0.16)$ and $(0.53,0.29,0.19)$ for MYYNOS, 
$(0.93,0.41,0.14)$ and $(0.79,0.29,0.15)$ for PP, and 
$(0.62,0.26,0.08)$ and $(0.45,0.21,0.17)$ for BSS. Note that 
the NLO results of BSS for different $\mu_{\rm soft}$ differ
qualitatively from all others. This is a consequence of a different 
treatment of the short-distance coefficient $C$ as explained above. 
The stability of the results from HT is mainly a consequence of
the use of a cutoff regularization scheme, which does not allow for
any momenta larger than the cutoff $\Lambda$ in the Green function of 
Eq.~(\ref{SchroedingerLO}), and of the fact that they solved
Eq.~(\ref{SchroedingerLO}) exactly rather than treating higher order
corrections perturbatively. 
All other groups use regularization
schemes that allow for infinitely large (i.e. relativistic) momenta in
Eq.~(\ref{SchroedingerLO}), in particular when their contributions do
not lead to ultraviolet divergences. The existence of this cutoff in
the result of HT implies that the meaning ``LO approximation'' is
modified. (See Ref.~\cite{Hoang2} for a discussion of the
cutoff-dependence of the results obtained by HT.) 
While LO approximation for all other groups means that 
all terms $v (\alpha_s/v)^n$ in the full QCD cross section are summed 
and no others are included, the LO approximation of HT contains 
cutoff-dependent terms that represent higher order short-distance
corrections. The difference between the results 
obtained by HT and the others indicates the size of these
higher order terms.   
Solving the Schr\"odinger equation~(\ref{SchroedingerLO}) exactly
rather than perturbatively, on the other hand, has the most
impact at NNLO, which can be seen from the NNLO scale variation in the
results from MYYNOS compared to the results from PP and BSS. The
results show that the resummation of the corrections of the NNLO
contributions in the Sch\"odinger equation~(\ref{SchroedingerLO}) to
all orders leads to a partial compensation of the large (fixed order)
NNLO corrections. 
However, we are not aware of any formal argument that the exact
solution of an approximate equation of motion in the framework of an
effective theory should {\it a priori} lead to a more reliable result
than the perturbative one. Which of the scale dependences provides a
more realistic estimate of yet higher order corrections can only be
answered when the full NNNLO corrections have been calculated.

The introduction of ``threshold  masses'' does not lead to a
reduction of the large NNLO normalisation
corrections (see the discussion in Sec.~\ref{sectionmasses}.) 
At the present stage, a final estimate for the normalisation
uncertainty of the total cross section at NNLO is difficult. In view
of the different behaviour of the NNLO corrections calculated by the
various groups, the variation of the normalisation with respect to
changes in $\mu_{\rm soft}$ seems not to be a reliable estimator. We
take the size of the NNLO correction to $R$ at the peak position at
$\mu_{\rm soft}=30$~GeV as an estimate for the current normalisation
uncertainty of the NNLO total cross section, which amounts to
about 20\%. This estimate is consistent with the variation of the
NNLO peak cross section with the used different calculational
methods by the various groups at $\mu_{\rm soft}=30$~GeV. Just
recently, some NNNLO corrections to the zero-distance wave function of
a (stable) toponium 1S state have been
determined~\cite{Kniehl1,Kniehl2}. Because the value of $R$ at the 
peak is proportional 
to the square of the toponium 1S wave function at the origin, these
corrections can be used as a consistency check for the error estimate
based on the NNLO corrections alone. In Ref.~\cite{Kniehl1} the
ultrasoft corrections (coming from the interactions of the top quarks
with dynamical gluons) were calculated, and in Ref.~\cite{Kniehl2} the
leading logarithmic contributions proportional to
$\ln^2(\alpha_s)$. Both contributions are below 10\%, which seems to 
support the error estimate of 20\% given above. However, a concrete
statement about the true size of the NNNLO corrections can only be
drawn once the full NNNLO corrections have been determined. 
In Ref.~\cite{condensate} non-perturbative corrections originating
from the gluon condensate have been calculated. These corrections
amount to less than a per cent in the normalisation and are negligible
compared to the current perturbative uncertainties.

Simulation studies~\cite{Peralta1} have shown that the normalisation
uncertainty does not seem to affect significantly the determination of
the top quark mass. However, it jeopardises the measurements of top
quark couplings from the threshold scan.

\section{Threshold Masses}
\label{sectionmasses}
The pole mass definition seems to be the natural choice to formulate
the non-relativistic effective theory that describes the $t\bar t$
dynamics close to threshold. The heavy quark pole mass is IR-finite
and gauge-invariant. In the pole mass scheme the equation of
motion for the non-relativistic $t\bar t$ pair has the simple
form of Eq.~(\ref{SchroedingerLO}), which is well known from
non-relativistic problems in QED. Intuition also seems to favour the
pole mass definition, because close to threshold the top quarks only
have a very small virtuality of order $M_t^2 v^2$. However, it is known
that the use of the pole mass can lead to (artificially) large high 
order corrections, because of its strong sensitivity to small
momenta~\cite{renormalon}. The results for the corrections to the peak
position obtained by all groups show that this is also the case for
the total 
$t\bar t$ production cross section. Technically, in the calculations
for the total cross section, the origin of the large corrections to
the peak position is the heavy quark potential (which is traditionally  
always given in the pole mass scheme). At large orders of
perturbation theory the potential causes large corrections from
momenta smaller than $M_t\alpha_s$, the relevant momentum scale for the
non-relativistic dynamics of the $t\bar t$
system~\cite{Aglietti1}. This can be 
visualised by considering the small momentum contribution to the
heavy quark potential in configuration space representation for
distances of the order of the inverse Bohr radius $1/M_t\alpha_s$:
\begin{eqnarray}
\bigg[\,V(r\approx 1/M_t\alpha_s)\,\bigg]^{\rm IR}
& \sim & 
\int\limits^{|\mbox{\scriptsize\boldmath $q$}|<\mu\ll M_t\alpha_s}
\frac{d^3\mbox{\boldmath $q$}}{(2\pi)^3}\,
\tilde V(\mbox{\boldmath $q$})\,
\exp(-i\,\mbox{\boldmath $q$}\mbox{\boldmath $r$})
\\[2mm] & \approx &
\int\limits^{|\mbox{\scriptsize\boldmath $q$}|<\mu\ll M_t\alpha_s}
\frac{d^3\mbox{\boldmath $q$}}{(2\pi)^3}\,
\tilde V_c(\mbox{\boldmath $q$}) \, + \, \ldots
\,.
\label{lowmomV}
\end{eqnarray}
Here, $\tilde V_c$ is the static potential in momentum space
representation. At large orders of perturbation theory the RHS of
Eq.~(\ref{lowmomV}) is dominated by $r$-independent corrections, which
grow asymptotically like $-\mu\,\alpha_s^n n!$. It has been
shown that the total static energy $2 M_t^{\rm pole} + V_c(r)$ does
not contain these large
corrections~\cite{Hoang3,Beneke2,Uraltsev1}\footnote{
In Refs.~\cite{V0const}, in the framework of potential models, it had
already been noted that the small momentum part of the static
heavy-quark--antiquark potential corresponds to a constant in the
potential in configuration space representation. This constant was
considered as an arbitrary and incalculable number universal to all
heavy-quark--antiquark systems, which would cancel for example in the
difference between the top and the bottom quark pole mass, up to 
mass-suppressed corrections. It was not realized, however, that the
corresponding ambiguity does not exist in the total static energy. 
}. 
Thus 
the large high order corrections can be avoided if a top quark mass
definition is adopted 
that does not contain the same strong sensitivity to small momenta as
the pole mass. Such masses are called ``short-distance''
masses. With a careful definition their ambiguity is parametrically of
order $\Lambda_{\rm QCD}^2/M_t$ or smaller. However, in the context of the
non-relativistic effective theory only those short-distance masses are
useful that differ from 
the pole mass by terms that are at most of the order of the non-relativistic 
energy of the top quarks in the $t\bar t$ system, i.e. of order
$M_t\alpha_s^2$. 
A difference that is parametrically larger than $M_t\alpha_s^2$
(such as $M_t\alpha_s$) would formally break the ``power counting'' of
the non-relativistic effective theory \cite{Beneke2}. This breakdown can be
visualised in the Schr\"odinger equation~(\ref{SchroedingerLO}), where
all terms are of order $M_t v^2\sim M_t\alpha_s^2$: expressing the
pole mass by a short-distance mass $m_t^{\rm sd}$ plus $\delta
m_t^{\rm sd}\equiv M_t-m_t^{\rm sd}\sim m_t^{\rm sd}\alpha_s$ would
make  $\delta m_t^{\rm sd}$ the dominant term in
Eq.~(\ref{SchroedingerLO}). From the formal point of view, this
excludes the $\overline{\mbox{MS}}$ mass from being a
useful threshold mass.

Three threshold mass parameters have been proposed so far:
Beneke suggested the ``potential-subtracted mass''
($m_t^{\rm PS}$)~\cite{Beneke2}, 
Hoang--Teubner suggested the ``1S mass''~\cite{Hoang2}, and  
Bigi {\it et al.} the ``kinetic mass''~\cite{Bigi2} (also called  
``low-scale running mass'' in some publications). The latter has
originally been devised to improve the perturbation series of
semileptonic B decay partial widths, but can be equally well applied
to heavy-quark--antiquark systems, because the large order behaviour
of the RHS of Eq.~(\ref{lowmomV}) is universal. The PS mass is defined by
\begin{eqnarray}
m_{t, \,\mu_f^{\rm PS}}^{\rm PS}
 & = &
M_t^{\rm pole} + \frac{1}{2}\,
\int\limits^{|\mbox{\scriptsize\boldmath $q$}|<\mu_f^{\rm PS}}
\frac{d^3\mbox{\boldmath $q$}}{(2\pi)^3}\,
\tilde V_c(\mbox{\boldmath $q$})
\nonumber \\[2mm] & = &
M_t^{\rm pole} - \frac{4}{3}\,\frac{\alpha_s}{\pi}\,\mu_f^{\rm PS}
\, + \, \ldots
\,,
\label{PSdef}
\end{eqnarray}
and can be regarded as the minimalistic way to eliminate the large
order corrections in Eq.~(\ref{lowmomV}).    
The 1S mass is defined as one half of the mass of the perturbative
contribution of a fictitious 
$n=1$, ${}^3S_1$ toponium bound state, assuming that the top quark is
a stable particle: 
\begin{eqnarray}
m_t^{\rm 1S} & = & \frac{1}{2}\,
\left[\,M_{\Upsilon_{t\bar t}(1S)}
\,\right]_{\rm pert}
\nonumber \\[2mm] & = &
M_t^{\rm pole} - \frac{2}{9}\,\alpha_s^2\,M_t^{\rm pole} 
\, + \, \ldots
\,.
\label{1Sdef}
\end{eqnarray}
The NNLO expression for the RHS of Eq.~(\ref{1Sdef}) was first 
calculated in Ref.~\cite{Pineda2}. 
The 1S scheme is motivated by the fact that twice the 1S mass is equal
to the peak of the total cross section up to corrections coming from
the finite top width. By construction, the 1S scheme strongly reduces
the correlation of the mass parameter to other theoretical parameters.
The kinetic mass is defined as
\begin{eqnarray}
m_{t,\,\mu_f^{\rm kin}}^{\rm kin}
& = &
M_t^{\rm pole} - \left[\bar\Lambda(\mu_f^{\rm kin})\right]_{\rm pert} 
- \left[\frac{\mu^2_\pi(\mu_f^{\rm kin})}{2 M_t^{\rm pole}}\right]_{\rm pert}
+\ldots
\nonumber \\[2mm] & = &
M_t^{\rm pole} - \frac{16}{9}\,\frac{\alpha_s}{\pi}\,\mu_f^{\rm kin}
\, + \, \ldots
\,,
\label{kindef}
\end{eqnarray}
where $\left[\bar\Lambda(\mu_f^{\rm kin})\right]_{\rm pert}$ and
$\left[\mu_\pi^2(\mu_f^{\rm kin})\right]_{\rm pert}$ are perturbative
evaluations of matrix elements of operators (defined in ``heavy quark
effective theory'', an effective theory widely employed in the theory
of B meson decays) that describe the difference between the pole and
the B meson mass. The two-loop contributions to the kinetic mass have
been calculated in Ref.~\cite{Czarnecki1}.
In the first line of Eq.~(\ref{kindef}) the ellipses
indicate matrix elements of higher dimension operators, which have not
been taken into account for this comparison.
In Eqs.~(\ref{PSdef}--\ref{kindef}) the respective
first order corrections have also been displayed. 
The PS and the kinetic masses depend on the scales $\mu_f^{\rm PS}$
and $\mu_f^{\rm kin}$, respectively. These scales are used as a cutoff
for the corresponding momentum integrations and cannot be chosen
parametrically larger than 
$M_t\alpha_s$  to preserve the non-relativistic power counting rules.  
For $\mu_f^{\rm PS}=\mu_f^{\rm kin}=0$ the PS and the kinetic masses
are equal to the pole mass. 
The 1S mass is cutoff-independent.

\begin{table}[tb]  
\vskip 7mm
\begin{center}
\begin{tabular}{|c||c|c|c|c||c|c|c|c|} \hline
$\overline m_t(\overline m_t)$ [GeV] &
  \multicolumn{4}{|c||}{$m_{t, 20\,\mbox{\scriptsize GeV}}^{\rm PS}$ [GeV]} &
  \multicolumn{4}{|c|}{$m_t^{\rm 1S}$ [GeV]} \\ \hline
& 1-loop & 2-loop & 3-loop & 4-loop 
& 1-loop & 2-loop & 3-loop & 4-loop \\ \hline
\multicolumn{9}{|c|}{$\alpha_s(M_z)=0.116$ 
          [$\alpha_s(\mbox{165 GeV})=0.1066$]} \\ \hline
160.00 & 166.36 & 167.49 & 167.75 & 167.82$^*$ 
       & 166.86 & 168.03 & 168.25 & -- \\ \hline
165.00 & 171.56 & 172.72 & 172.99 & 173.06$^*$ 
       & 172.05 & 173.24 & 173.46 & -- \\ \hline
170.00 & 176.76 & 177.96 & 178.23 & 178.31$^*$ 
       & 177.23 & 178.45 & 178.68 & -- \\ \hline
\multicolumn{9}{|c|}{$\alpha_s(M_z)=0.119$ 
          [$\alpha_s(\mbox{165 GeV})=0.1091$]} \\ \hline
160.00 & 166.51 & 167.69 & 167.97 & 168.05$^*$
       & 167.02 & 168.23 & 168.46 & -- \\ \hline
165.00 & 171.72 & 172.93 & 173.22 & 173.30$^*$
       & 172.21 & 173.45 & 173.68 & -- \\ \hline
170.00 & 176.92 & 178.17 & 178.47 & 178.55$^*$
       & 177.39 & 178.67 & 178.91 & -- \\ \hline
\multicolumn{9}{|c|}{$\alpha_s(M_z)=0.122$ 
          [$\alpha_s(\mbox{165 GeV})=0.1117$]} \\ \hline
160.00 & 166.66 & 167.90 & 168.20 & 168.28$^*$
       & 167.17 & 168.43 & 168.68 & -- \\ \hline
165.00 & 171.87 & 173.15 & 173.45 & 173.54$^*$
       & 172.36 & 173.65 & 173.90 & -- \\ \hline
170.00 & 177.08 & 178.39 & 178.70 & 178.80$^*$
       & 177.55 & 178.88 & 179.13 & -- \\ \hline
\end{tabular}
\caption{\label{tabps1s} 
Top quark PS and 1S mass values for a given value of the top quark
$\overline{\mbox{MS}}$ mass $\overline m_t$ at the scale $\overline
m_t$ for $\alpha_s(M_Z)=0.116$, $0.119$ and $0.121$.
Large-$\beta_0$ estimates are indicated by a star.
}
\end{center}
\vskip 3mm
\end{table}
\begin{table}[tb]  
\vskip 7mm
\begin{center}
\begin{tabular}{|c||c|c|c|c||c|c|c|c|} \hline
$\overline m_t(\overline m_t)$ [GeV] &
  \multicolumn{4}{|c||}{$m_{t, 15\,\mbox{\scriptsize GeV}}^{\rm kin}$ [GeV]} &
  \multicolumn{4}{|c|}{$M^{\rm pole}_t$ [GeV]} \\ \hline
& 1-loop & 2-loop & 3-loop & 4-loop 
& 1-loop & 2-loop & 3-loop & 4-loop \\ \hline
\multicolumn{9}{|c|}{$\alpha_s(M_z)=0.116$ 
          [$\alpha_s(\mbox{165 GeV})=0.1066$]} \\ \hline
160.00 & 166.33 & 167.38 & 167.63$^*$ & --  
       & 167.27 & 168.80 & 169.28 & 169.50$^*$ \\ \hline
165.00 & 171.53 & 172.62 & 172.87$^*$ & --  
       & 172.47 & 174.03 & 174.52 & 174.74$^*$ \\ \hline
170.00 & 176.73 & 177.85 & 178.12$^*$ & --  
       & 177.66 & 179.26 & 179.75 & 179.98$^*$ \\ \hline
\multicolumn{9}{|c|}{$\alpha_s(M_z)=0.119$ 
          [$\alpha_s(\mbox{165 GeV})=0.1091$]} \\ \hline
160.00 & 166.48 & 167.58 & 167.85$^*$ & -- 
       & 167.44 & 169.05 & 169.56 & 169.80$^*$ \\ \hline
165.00 & 171.68 & 172.83 & 173.10$^*$ & --
       & 172.64 & 174.28 & 174.80 & 175.05$^*$ \\ \hline
170.00 & 176.89 & 178.07 & 178.35$^*$ & --
       & 177.84 & 179.52 & 180.05 & 180.30$^*$ \\ \hline
\multicolumn{9}{|c|}{$\alpha_s(M_z)=0.122$ 
          [$\alpha_s(\mbox{165 GeV})=0.1117$]} \\ \hline
160.00 & 166.63 & 167.79 & 168.07$^*$ & --
       & 167.61 & 169.29 & 169.84 & 170.11$^*$ \\ \hline
165.00 & 171.84 & 173.03 & 173.32$^*$ & --
       & 172.82 & 174.53 & 175.09 & 175.36$^*$ \\ \hline
170.00 & 177.05 & 178.28 & 178.58$^*$ & --
       & 178.02 & 179.77 & 180.34 & 180.61$^*$ \\ \hline
\end{tabular}
\caption{\label{tabkinpole} 
Top quark kinetic and pole mass values for a given value of the top
quark $\overline{\mbox{MS}}$ mass $\overline m_t$ at the scale
$\overline m_t$ for $\alpha_s(M_Z)=0.116$, $0.119$ and
$0.122$. Large-$\beta_0$ estimates are indicated by a star. 
}
\end{center}
\vskip 3mm
\end{table}
\begin{figure}[t!] 
\begin{center}
\leavevmode
\epsfxsize=3.8cm
\leavevmode
\epsffile[220 580 420 710]{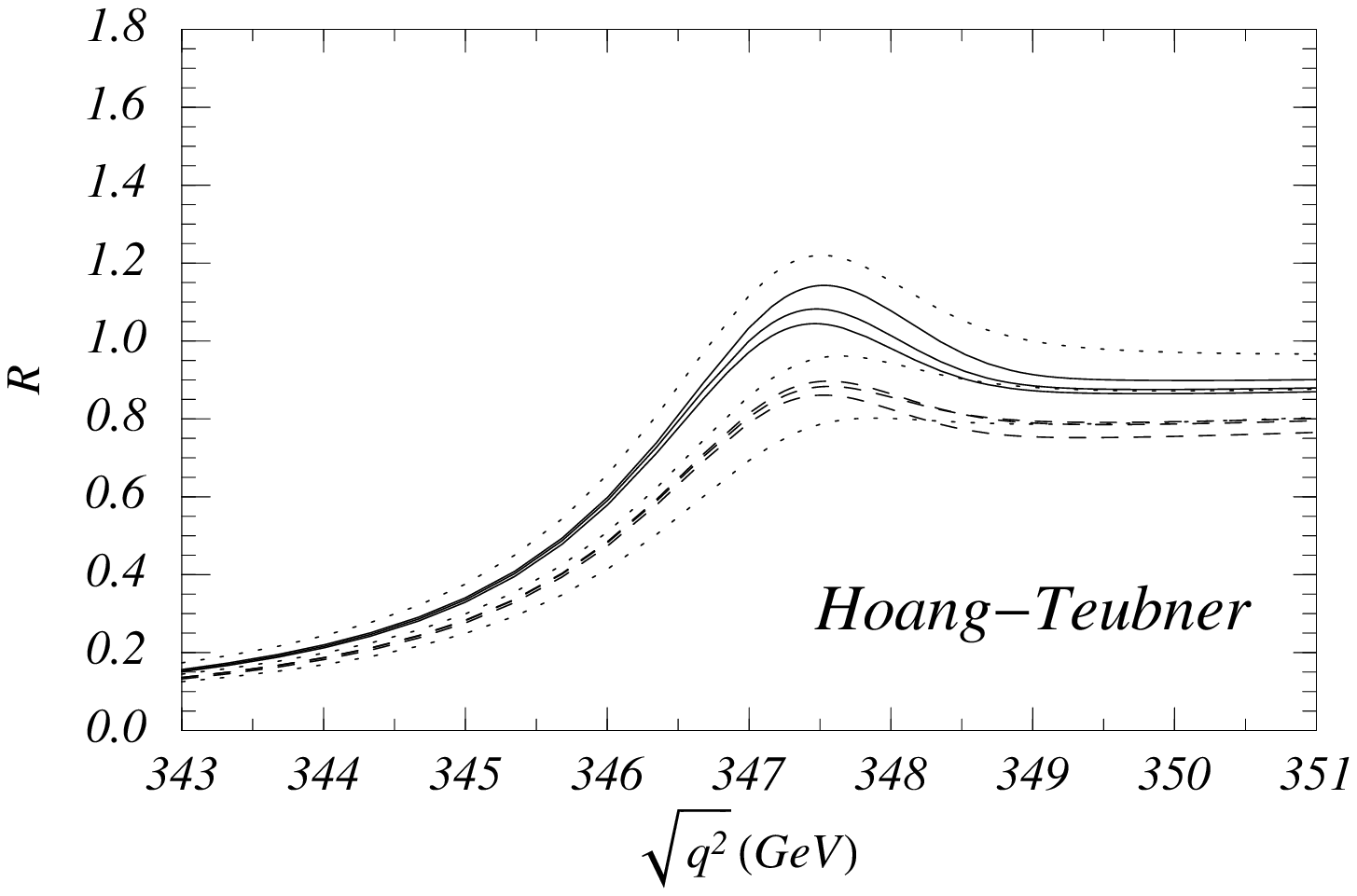}
\hspace{4.1cm}
\epsfxsize=3.8cm
\leavevmode
\epsffile[220 580 420 710]{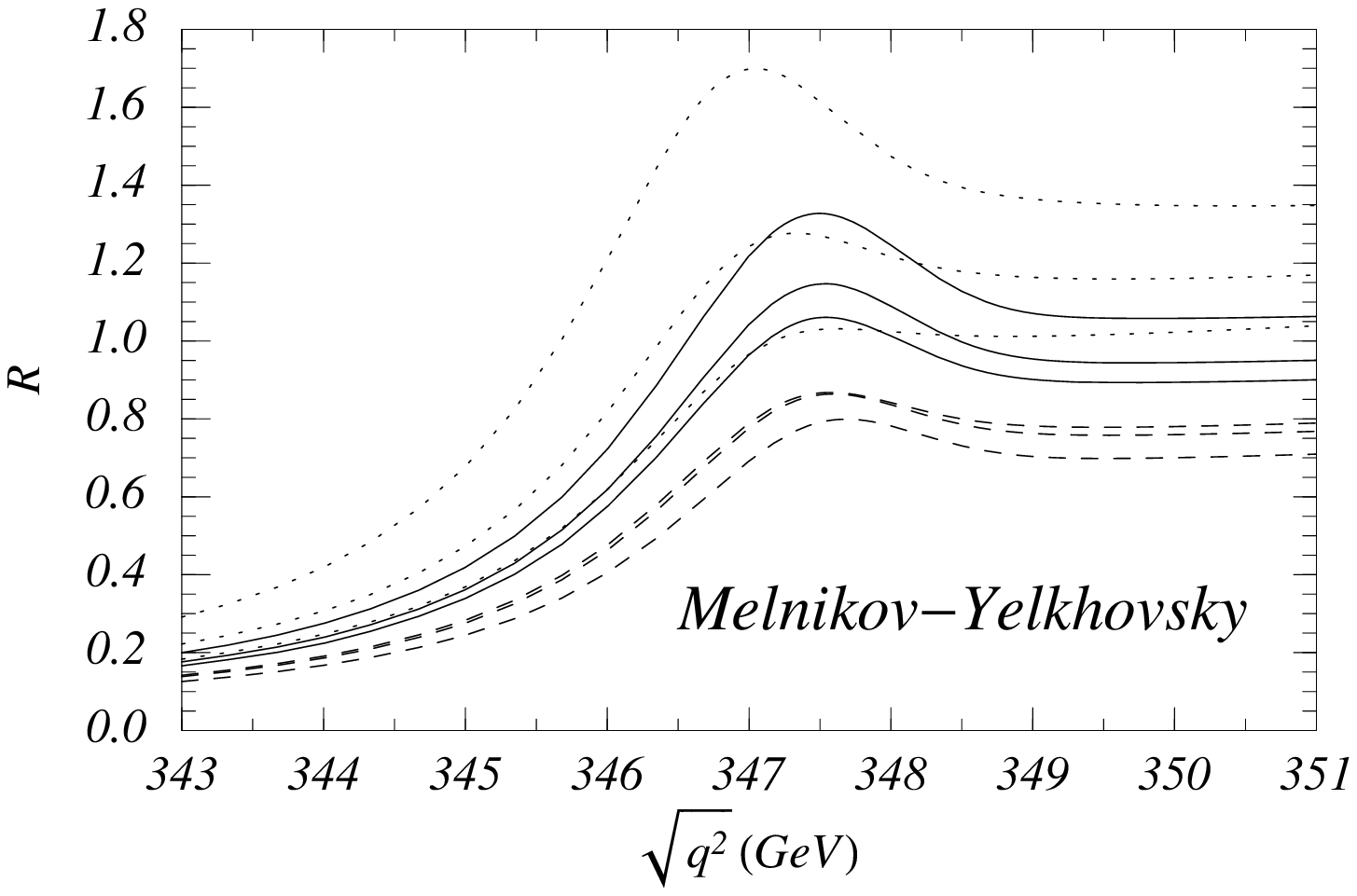}
\vskip 3cm
\leavevmode
\epsfxsize=3.8cm
\leavevmode
\epsffile[220 580 420 710]{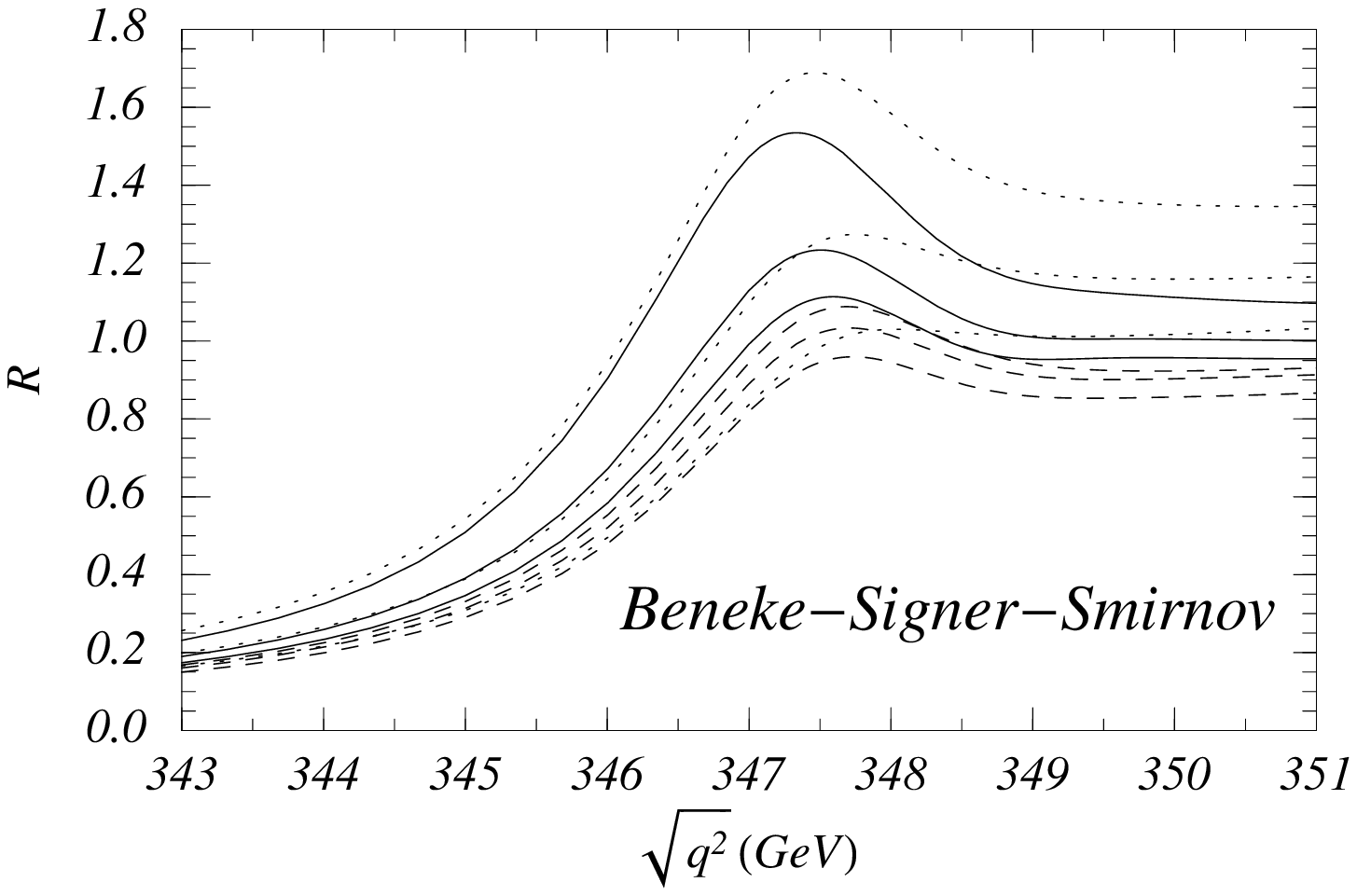}
\hspace{4.1cm}
\epsfxsize=3.8cm
\leavevmode
\epsffile[220 580 420 710]{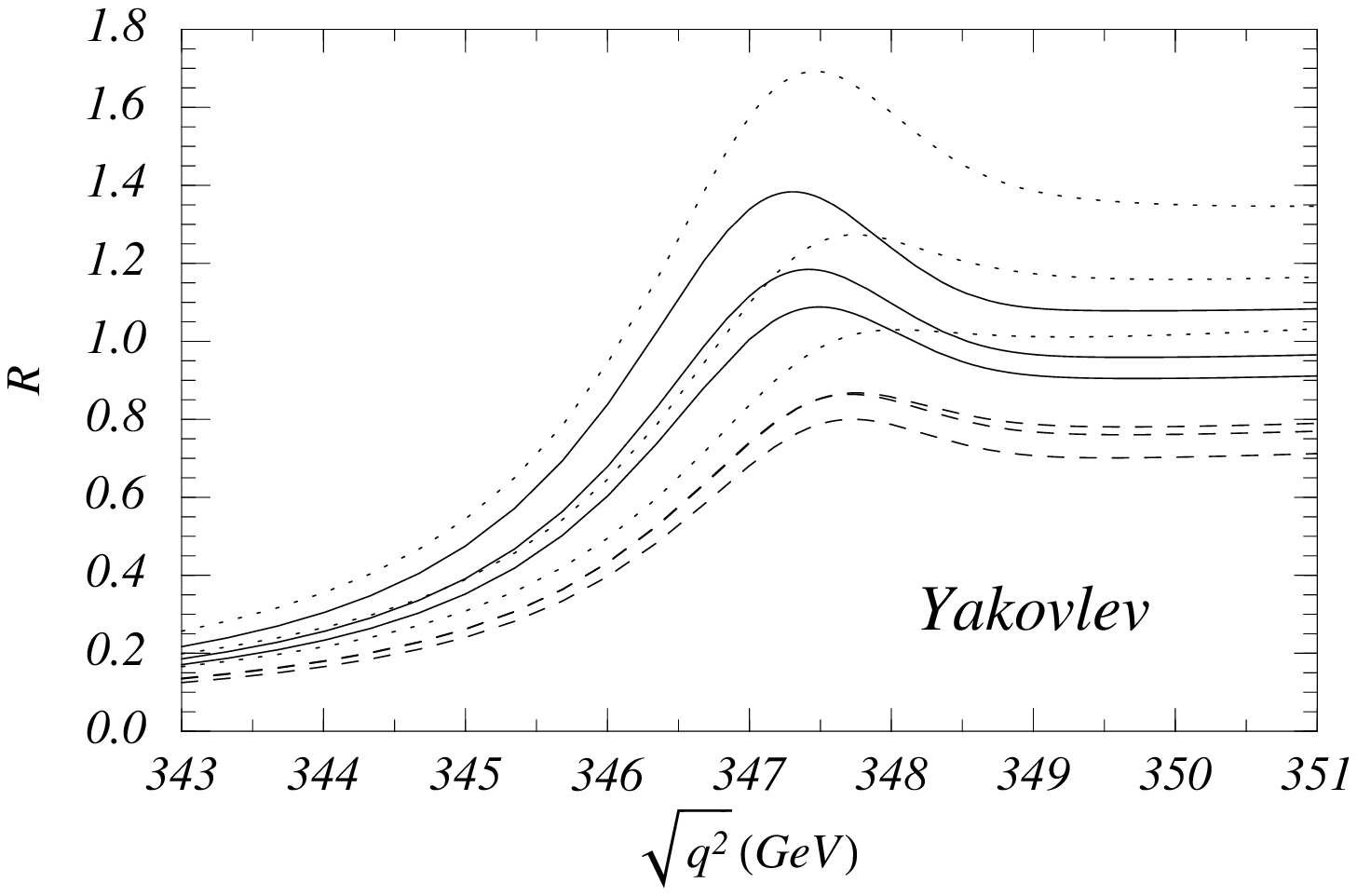}
\vskip  2.5cm
 \caption{\label{figthresholdmasses}
The total normalised photon-induced $t\bar t$ cross section  $R$ at
the LC versus the c.m. energy in the threshold regime at LO (dotted
curves), NLO (dashed) and NNLO (solid). Hoang--Teubner used the 1S mass
scheme with $m_t^{1S}=173.68$~GeV, Melnikov--Yelkhovsky the kinetic
mass at $15$~GeV with 
$m_{t, 15\,\mbox{\scriptsize GeV}}^{\rm kin}=173.10$~GeV, and 
Beneke--Signer--Smirnov and Yakovlev the PS mass at $20$~GeV with
$m_{t, 20\,\mbox{\scriptsize GeV}}^{\rm PS}=173.30$~GeV.
The plots have been generated from results provided by the groups
Hoang--Teubner (HT), Melnikov--Yelkhovsky (MY) and Beneke--Signer--Smirnov
(BSS) and Yakovlev. 
}
 \end{center}
\end{figure}
\begin{table}[t] 
\vskip 7mm
\begin{center}
\begin{tabular}{|c||c|c|c||c|c|c||c|c|c|} \hline
Order & 
\multicolumn{3}{|c||}{LO} 
   & \multicolumn{3}{|c||}{NLO}
   & \multicolumn{3}{|c|}{NNLO} \\ \hline
$\mu_{\rm soft}$[GeV] & $15$ & $30$ & $60$ 
                      & $15$ & $30$ & $60$  
                      & $15$ & $30$ & $60$  \\ \hline
\raisebox{-1.6ex}[1.6ex]{HT ($m_t^{1S}$)}
  & $1.22$ & $0.96$ & $0.80$
  & $0.86$ & $0.90$ & $0.88$
  & $1.14$ & $1.08$ & $1.04$ \\[-1mm]
  & $347.51$ & $347.65$ & $347.88$
  & $347.53$ & $347.56$ & $347.57$
  & $347.53$ & $347.48$ & $347.46$ \\\hline
\raisebox{-1.6ex}[1.6ex]{MY 
($m^{\rm kin}_{t, 15~\mbox{\scriptsize GeV}}$)}
  & $1.70$ & $1.28$ & $1.03$
  & $0.80$ & $0.86$ & $0.87$ 
  & $1.33$ & $1.15$ & $1.06$ \\[-1mm]
  & $347.05$ & $347.33$ & $347.63$
  & $347.69$ & $347.59$ & $347.58$
  & $347.50$ & $347.54$ & $347.54$ \\\hline
\raisebox{-1.6ex}[1.6ex]{BSS 
($m^{\rm PS}_{t, 20~\mbox{\scriptsize GeV}}$)}
  & $1.69$ & $1.27$ & $1.03$ 
  & $1.09$ & $1.03$ & $0.96$
  & $1.53$ & $1.23$ & $1.11$ \\[-1mm]
  & $347.47$ & $347.74$ & $348.05$
  & $347.69$ & $347.71$ & $347.73$ 
  & $347.33$ & $347.51$ & $347.59$ \\\hline
\end{tabular}
\caption{\label{tabthresholdmasses}
The values of $R$ (upper numbers) at the respective peak position
(lower numbers in units of GeV) at LO, NLO and NNLO.
Hoang-Teubner used the 1S mass
scheme with $m_t^{1S}=173.68$~GeV, Melnikov--Yelkhovsky the kinetic
mass at $15$~GeV with 
$m_{t, 15\,\mbox{\scriptsize GeV}}^{\rm kin}=173.10$~GeV, and
Beneke-Signer-Smirnov the PS mass at $20$~GeV with
$m_{t, 20\,\mbox{\scriptsize GeV}}^{\rm PS}=173.30$~GeV.
The values have been determined
from results provided by the groups Hoang--Teubner (HT),
Melnikov--Yelkhovsky (MY) and Beneke--Signer--Smirnov (BSS).  }
\end{center}
\vskip 3mm
\end{table}

The three threshold masses eliminate the large higher order
corrections to the peak position mentioned above. In
addition, they can reduce the correlation of the
peak position to the value of the strong coupling and theoretical
parameters, such as the renormalisation scale $\mu_{\rm soft}$. For
the 1S mass this is achieved automatically; for the PS and the kinetic
mass this is achieved by setting $\mu_f$ to a value of order
$M_t\alpha_s\approx 15$--$20$~GeV. (Choosing $\mu_f$ much smaller than
$M_t\alpha_s$ also eliminates the large corrections at high orders,
but does not reduce the correlation to $\mu_{\rm soft}$ and $\alpha_s$
in a significant way~\cite{Nagano1}.) 
In Tables~\ref{tabps1s} and \ref{tabkinpole} numerical values of the
top quark PS mass for $\mu_f^{\rm PS}=20$~GeV, the 1S mass, and the
kinetic mass for $\mu_f^{\rm kin}=15$~GeV are given, taking the
$\overline{\mbox{MS}}$ mass 
$\overline m_t(\overline m_t)=160,165,170$~GeV as a reference point
and using $\alpha_s(165~\mbox{GeV})=0.1066,0.1091,0.1117$. As a
comparison, also the 
corresponding values for the top quark pole mass have been displayed in
Table~\ref{tabkinpole}. The
knowledge of the two- \cite{Gray1} and three-loop
corrections~\cite{Chetyrkin1,Ritbergen1}\footnote{
To obtain the numerical values given in Tables~\ref{tabps1s} and
\ref{tabkinpole} we used Eq.~(6) of the first publication of
Ref.~\cite{Chetyrkin1}.
} 
in the relation 
between the pole and the $\overline{\mbox{MS}}$ mass are required to
obtain the two- and three-loop values of the PS, 1S and kinetic mass.
For the PS, 1S and pole masses the
relation to the $\overline{\mbox{MS}}$ mass is known at three
loops and for the kinetic mass at two loops. For
the PS mass the four-loop contributions in the ``large-$\beta_0$''
limit have been derived from the expression for the  
Borel transform of the static potential~\cite{Aglietti1} and of the 
difference between the pole and the $\overline{\mbox{MS}}$
mass~\cite{polemsbar}. The same information is in principle sufficient
to determine the four-loop ``large-$\beta_0$'' correction in the
difference between the 1S and the $\overline{\mbox{MS}}$ mass. 
For the kinetic mass the three-loop
contributions in the ``large-$\beta_0$'' limit have been determined 
in~\cite{Czarnecki1}. (The three-loop ``large-$\beta_0$'' corrections
in the relation between the kinetic and the $\overline{\mbox{MS}}$
mass have been obtained using Eq.~(21) in the preprint version of
Ref.~\cite{Czarnecki1}.) 
In Tables~\ref{tabps1s} and \ref{tabkinpole} the large-$\beta_0$
corrections are indicated by a star. We emphasise that the numbers
shown in these tables do not contain any
electroweak corrections. The latter can amount to shifts 
at the $1$ GeV level~\cite{Hempfling1}. 

The numbers displayed in the
tables show an excellent convergence of the perturbative relation
between the threshold masses and the $\overline{\mbox{MS}}$ mass.
For $\alpha_s(M_Z)=0.119$ and $\overline m_t(\overline m_t)=165$~GeV
the one-, two-, three- and the available four-loop large-$\beta_0$
corrections for the threshold masses are $6.7$--$7.2$, $1.2$,
$0.2$--$0.3$ and $0.1$~GeV, 
respectively. The corresponding corrections in the relation between
the pole mass and $\overline m_t(\overline m_t)$ read
$7.6$, $1.6$, $0.5$ and $0.3$~GeV. For the pole mass the three-loop
corrections are about a factor two and the four-loop large-$\beta_0$
corrections a factor three larger. This behaviour is
caused by the infrared-sensitivity of the pole mass and corresponds to
the ambiguity of the pole mass of order $\Lambda_{\rm QCD}$. 
The numbers displayed in the tables also show that a
shift in $\alpha_s(M_Z)$ by $0.001$ corresponds to a shift of about
$70$~MeV in the threshold masses. 
 
In the framework of the Linear Collider Workshop 
several presentations were given by M.~Beneke, A.~H.~Hoang,
K.~Melnikov, Y.~Sumino, T.~Teubner and O.~Yakovlev
of, in part, preliminary results for the cross
section using threshold masses. The following
discussion provides a summary of these results, choosing one
representative example for each of the three threshold mass
definitions. 
The results for the discussion have been provided by HT in the 1S
scheme  for $m_t^{1S}=173.68$~GeV, by Melnikov--Yelkhovsky (MY) in the
kinetic mass scheme for 
$m^{\rm kin}_{t, 15\,\mbox{\scriptsize GeV}}=173.10$~GeV, and by BSS
in the PS mass scheme for   
$m^{\rm PS}_{t, 20\,\mbox{\scriptsize GeV}}=173.30$~GeV. The numerical
values of the respective threshold masses are the known highest order
entries in Tables~\ref{tabps1s} and \ref{tabkinpole} for common      
$\alpha_s(M_Z)=0.119$ and $\overline m_t(\overline m_t)=165$~GeV.
HT and BSS used the codes developed for Refs.~\cite{Hoang2} and
\cite{Beneke1}, respectively. Yakovlev and NOS have also provided
results in the PS mass scheme. Their results are in qualitative
agreement with those of BSS. In Figs.~\ref{figthresholdmasses} the total
normalised photon-induced cross section $R$ is displayed at LO
(dotted lines), NLO (dashed lines) and NNLO (solid lines) using 
the three threshold masses mentioned above for $\alpha_s(M_Z)=0.119$
and $\mu_{\rm soft}=15$, $30$, $60$~GeV, and ignoring the effects of
beamstrahlung and initial state radiation.
The values of $R$ (upper number) at the respective peak position
(lower number in units of GeV) are given in
Table~\ref{tabthresholdmasses}.

The threshold masses have been implemented employing the
non-relativistic power-counting rules for the perturbative series
describing the difference between threshold and pole mass.
This means that for all threshold masses the one-loop corrections
displayed in Eqs.~(\ref{PSdef}--\ref{kindef}) have been treated as LO
in the non-relativistic expansion, the two-loop corrections as NLO and
so on.  The results in Figs.~\ref{figthresholdmasses} and
Table~\ref{tabthresholdmasses} show that the peak positions obtained
with different threshold masses converge when higher orders are 
included. This is a consequence of the fact that the numerical values
of all the threshold masses have been determined from $\overline
m_t(\overline m_t)=165$~GeV as a reference value.  
Compared to the results in the pole mass
scheme displayed in Sec.~\ref{sectionresults}, the results in
Figs.~\ref{figthresholdmasses} and Table~\ref{tabthresholdmasses} 
show an improved stability of the peak position with respect to the
size of higher order corrections, and with respect to the sensitivity
to changes in $\mu_{\rm soft}$.
For $\mu_{\rm soft}=15/30/60$~GeV, the NLO (NNLO) shifts of the peak
position are
$20/-90/-310$~MeV ($0/-80/-110$~MeV) for HT in the 1S scheme,
$640/260/-50$~MeV ($-190/-50/-40$~MeV) for MY in the kinetic mass
scheme and
$220/-30/-330$~MeV ($-360/-200/-140$~MeV) for BSS in the PS scheme.
The lack of convergence that can be observed for some numbers
is not an indication of large unknown higher order corrections, but a
consequence of the fact that the threshold masses that have been used
for the analysis partially lead to NLO shifts that are much smaller
than the parametric accuracy that can be achieved at the NLO level. 
A better quantification is obtained by comparing the shift 
from LO directly to NNLO with the corresponding shift, when the cross
section is plotted for fixed pole mass as in Figs.~\ref{figpole}. 
The variation of the peak position when $\mu_{\rm soft}$ is varied
between $15$ and $60$~GeV  at LO/NLO/NNLO is 
$370/40/70$~MeV for HT in the 1S scheme,
$580/110/40$~MeV for MY in the kinetic mass scheme and
$580/40/260$~MeV for BSS in the PS scheme.
The scale variation of the peak position at NNLO obtained by BSS in
the PS mass scheme is by a factor of 4--5 larger than the
corresponding variation obtained by HT in the 1S and by MY in the 
kinetic mass scheme. The fact that the stability of the
results in the PS mass scheme is worse than in the 1S and the
kinetic mass scheme might originate from the fact that the
difference between the PS and the pole mass contains only corrections
from the static potential. The 1S and the kinetic
mass contain additional corrections, which are subleading in the
non-relativistic velocity counting. However, it should be noted that 
the stability of the peak position for a fixed threshold mass is not 
necessarily a useful quantification of the theoretical error.  
A mass definition that would be equal to the peak
position, for example, would lead to no variation at all. 
The shifts of the peak position quoted above should therefore be
considered in conjunction with the variation of the threshold mass
values with the order of perturbation theory  
given in Tables~\ref{tabps1s} and~\ref{tabkinpole}. 
       
From the size of the NNLO corrections to the peak positions and 
from the scale variation of the peak positions at NNLO, we estimate
that the current theoretical uncertainty of a determination of the
threshold masses from the peak position based on the NNLO calculations
is about $100$~MeV. (The uncertainty coming from the different
calculational methods used by the various groups has been estimated in
Sec.~\ref{sectionresults} and is included in this number.) 
In Refs.~\cite{Brambilla1,Kniehl1,Kniehl2} the ultrasoft corrections
and the leading logarithmic contributions proportional to
$\ln\alpha_s$ were calculated for the mass of a fictitious toponium 1S
state at NNNLO. Up to corrections coming from the top width
(and other electroweak corrections), the toponium 1S mass is equal to
the location of the peak of $R$. As for the normalisation, these
corrections can be used as a consistency check for the error
estimate of the top mass extraction based on the NNLO corrections
alone. Both types of corrections amount to about $200$~MeV,
corresponding to a shift of $100$~MeV in the top quark mass. This
supports the error estimate based on the NNLO calculations alone.   
However, as for the case of the normalisation, a concrete statement
about the true size of the NNNLO corrections can only be drawn once
the full NNNLO corrections have been determined. 
Taking into account four-loop corrections in the
relation between the threshold masses and the $\overline{\mbox{MS}}$
mass, $\overline m_t(\overline m_t)$ can be determined with comparable
precision for an uncertainty in $\alpha_s(M_Z)$ of around $0.001$--$0.002$.
Realistic simulation studies~\cite{Peralta1} have shown that these
conclusions remain valid if beam effects from initial state radiation
or beamstrahlung are taken into account.  

The threshold masses do not lead to an improvement of the stability in
the normalisation of $R$ because their main effect is to redefine the
binding energy of the $t\bar t$ system. An energy shift leaves the
wave function of the $t\bar t$ system unaffected and, therefore,
cannot affect higher order corrections to the normalisation.
\section{Summary and Open Issues}
\label{sectionopen}
In this article the results for the NNLO
QCD calculations for the total photon-mediated $t\bar t$ production
cross section obtained by different groups have been compared. A
detailed assessment 
of the dependence of the individual results on the calculational
techniques, the intermediate regularization prescriptions and the
treatment of higher order corrections has been carried out. As far as
the determination of the top quark mass from the position of the peak
in the total cross section is concerned, the uncertainty caused by
the use of different methods is around 50--80~MeV. Using the top quark
pole mass to parameterise the total cross section, the latter
uncertainty is negligible with respect to the perturbative uncertainty in
an extraction of the pole mass parameter, which is estimated to be
around 300~MeV. This estimate matches formal arguments based on the
analysis of the large order behaviour of perturbation theory, which
state that the pole mass cannot be determined to a precision better
than ${\cal{O}}(\Lambda_{\rm QCD})$. Using so-called ``threshold masses'',
which lead to a much better high order behaviour and which preserve
the non-relativistic velocity counting, the uncertainty in the mass
extraction is around 100~MeV. The top quark $\overline{\mbox{MS}}$ mass
$\overline m_t(\overline m_t)$ can be determined with comparable
precision if $\alpha_s(M_Z)$ is known with an uncertainty of
$0.001$--$0.002$. (An uncertainty in $\alpha_s(M_Z)$ of $0.001$
corresponds to an uncertainty of 70~MeV in 
$\overline m_t(\overline m_t)$.) 
Realistic simulation studies have shown that these conclusions remain
valid if realistic beam effects are taken into account.
For the normalisation of the total cross section, we find that the NNLO
corrections are much larger than indicated by the renormalisation
scale dependence of the NLO results. The normalisation at NNLO also
has a considerable dependence on calculational methods and the
renormalisation scale $\mu_{\rm soft}$. We estimate the uncertainty of
the normalisation of the NNLO cross section as around 20\%, which
seems to jeopardise accurate measurements of top 
quark couplings or the total top quark width. The calculation of NNNLO
corrections will be mandatory to reduce the current uncertainties in the
normalisation of the total cross section. At present, the most
difficult parts of a complete NNNLO QCD calculation of the total cross
section seem to be the three-loop corrections to the static potential
and the short-distance coefficient $C$. Another way to get information
about the size of higher order corrections is to resum logarithmic
contributions to all orders in perturbation theory using the
renormalisation group evolution of the operators in the
non-relativistic effective theory for $t\bar t$ close to
threshold. First attempts at carrying out such a resummation
consistently have been made in Refs.~\cite{Beneke1,Manohar1}. 

Except for the NNLO calculations of the top three momentum distribution
in Refs.~\cite{Nagano1,Hoang2},
practically nothing is known about the size of NNLO corrections to
differential observables.  In view of the large NNLO QCD corrections
to the total cross section, calculations of the complete NNLO
corrections to differential observables, such as the top momentum
distribution, the forward--backward asymmetry and certain leptonic
spectra, are needed to obtain realistic estimates of the theoretical
uncertainties. 
A first step toward this aim is the development of a
consistent and systematic approach to account for electroweak
effects. The present calculations of the 
total $t\bar t$ cross section at NNLO have only taken into account 
relativistic corrections in the framework of QCD. So far, 
electroweak effects have been taken into account, mainly by
shifting the energy in the Schr\"odinger
equation~(\ref{SchroedingerLO}) into the upper complex plane by
$i\Gamma_t$. This treatment accounts for all electroweak  
effects at LO in the non-relativistic expansion. 
At NLO electroweak effects lead to final state interactions
originating from the exchange of gluons between the top quarks
and their decay products. These corrections are called
``non-factorisable'' (or ``rescattering'') corrections because they
can in general not be unambiguously considered as being either
corrections to $t\bar t$ production or to top quark decay. For the
total cross section it has been shown that the non-factorisable
corrections cancel at NLO and that the net effect of the electroweak
corrections reduces to shifting the c.m. energy by
$i\,\Gamma_t$~\cite{NLOnonfactortot}. For a number of differential
observables, such as the top quark momentum distribution and the
energy spectrum of leptons originating from the decay of a W-boson,
NLO non-factorisable corrections have been
calculated~\cite{NLOnonfactordist} and shown to be 
of order 10\%, as can be expected for ${\cal{O}}(\alpha_s)$
corrections. No consistent and systematic prescription to 
implement electroweak effects at NNLO has been developed yet, and
practically nothing is known about the size of the 
non-factorisable corrections beyond NLO. 

\par
\vspace{.5cm}
\section*{Acknowledgements}
This synopsis was initiated at the Mini-Workshop on Physics at the
$t\bar t$ Threshold, organised by J.~H.~K\"uhn and held at the
Institute for Theoretical Particle Physics, Karlsruhe, Germany, on
December 20, 1998. The participants of the workshop were A.~H.~Hoang,
J.~H.~K\"uhn, K.~Melnikov, T.~Teubner and O.~Yakovlev.
 
M. Beneke and A. H. Hoang are supported in part by the EU Fourth
Framework Program ``Training and Mobility of Researchers'', Network
``Quantum Chromodynamics and Deep Structure of Elementary Particles'',
contract FMRX-CT98-0194 (DG12-MIHT).
K. Melnikov is supported by the US Department of Energy
(DOE) under contract DE-AC03-76SF00515.  
T.~Nagano, A.~Ota and Y.~Sumino are supported in part by the
Japan-German Cooperative Science Promotion Program.
A.~A.~Pivovarov is partially supported by the Volkswagen Foundation
under contract No.~I/73611 and the Russian Fund for Basic Research
under contract No.~99-01-00091.
The work of V.A. Smirnov is partially supported by the Volkswagen
Foundation under contract No.~I/73611. 
O.~Yakovlev acknowledges partial support from the US Department of
Energy (DOE) and the Bundesministerium f\"ur Bildung und Forschung 
(BMFT) under contract 05-HT9WWA-9. A. Penin is supported by the
Bundesministerium f\"ur Bildung und Forschung under contract 
No.~05~HT9GUA 3 and by the Volkswagen Foundation. A.~Yelkhovsky is
partially supported by the Russian Ministry of Higher Education.

We thank W.~Bernreuther and J.~H.~K\"uhn as conveners of the Linear
Collider Top Quark Working Group, and J.~H.~K\"uhn for his comments
to the manuscript.                                              
\vspace{1.0cm}
%
\sloppy
\raggedright
\def\app#1#2#3{{\it Act. Phys. Pol. }{\bf B #1} (#2) #3}
\def\apa#1#2#3{{\it Act. Phys. Austr.}{\bf #1} (#2) #3}
\def\lhc{Proc. LHC Workshop, CERN 90-10}
\def\npb#1#2#3{{\it Nucl. Phys. }{\bf B #1} (#2) #3}
\def\npps#1#2#3{{\it Nucl. Phys. Proc. Suppl. }{\bf #1} (#2) #3}
\def\nP#1#2#3{{\it Nucl. Phys. }{\bf #1} (#2) #3}
\def\plb#1#2#3{{\it Phys. Lett. }{\bf B #1} (#2) #3}
\def\prd#1#2#3{{\it Phys. Rev. }{\bf D #1} (#2) #3}
\def\pra#1#2#3{{\it Phys. Rev. }{\bf A #1} (#2) #3}
\def\pR#1#2#3{{\it Phys. Rev. }{\bf #1} (#2) #3}
\def\prl#1#2#3{{\it Phys. Rev. Lett. }{\bf #1} (#2) #3}
\def\prc#1#2#3{{\it Phys. Reports }{\bf #1} (#2) #3}
\def\cpc#1#2#3{{\it Comp. Phys. Commun. }{\bf #1} (#2) #3}
\def\nim#1#2#3{{\it Nucl. Inst. Meth. }{\bf #1} (#2) #3}
\def\pr#1#2#3{{\it Phys. Reports }{\bf #1} (#2) #3}
\def\sovnp#1#2#3{{\it Sov. J. Nucl. Phys. }{\bf #1} (#2) #3}
\def\sovpJ#1#2#3{{\it Sov. Phys. LETP }{\bf #1} (#2) #3}
\def\jl#1#2#3{{\it JETP Lett. }{\bf #1} (#2) #3}
\def\jet#1#2#3{{\it JETP Lett. }{\bf #1} (#2) #3}
\def\zpc#1#2#3{{\it Z. Phys. }{\bf C #1} (#2) #3}
\def\ptp#1#2#3{{\it Prog.~Theor.~Phys.~}{\bf #1} (#2) #3}
\def\nca#1#2#3{{\it Nuovo~Cim.~}{\bf #1A} (#2) #3}
\def\ap#1#2#3{{\it Ann. Phys. }{\bf #1} (#2) #3}
\def\hpa#1#2#3{{\it Helv. Phys. Acta }{\bf #1} (#2) #3}
\def\ijmpA#1#2#3{{\it Int. J. Mod. Phys. }{\bf A #1} (#2) #3}
\def\ZETF#1#2#3{{\it Zh. Eksp. Teor. Fiz. }{\bf #1} (#2) #3}
\def\jmp#1#2#3{{\it J. Math. Phys. }{\bf #1} (#2) #3}
\def\yf#1#2#3{{\it Yad. Fiz. }{\bf #1} (#2) #3}
\def\ufn#1#2#3{{\it Usp. Fiz. Nauk }{\bf #1} (#2) #3}
\def\spu#1#2#3{{\it Sov. Phys. Usp.}{\bf #1} (#2) #3}
\def\epjc#1#2#3{{\it Eur. Phys. J. }{\bf C #1} (#2) #3}

\end{document}